\documentclass[10pt,superscriptaddress,twocolumn,amsmath,amssymb,aps,prl,showpacs]{revtex4-1}
\usepackage{mathrsfs}
\usepackage{graphicx}
\usepackage{dcolumn}
\usepackage{bm}
\usepackage{amssymb}
\usepackage{amsmath}
\usepackage{paralist}
\usepackage{float}
\usepackage{mdframed}
\usepackage{booktabs}
\usepackage{mathtools} 
\usepackage{graphicx}
\usepackage[colorlinks,linkcolor=blue,anchorcolor=blue,citecolor=green]{hyperref} 
\usepackage{cleveref}
\usepackage{url}
\usepackage{color}

\newcommand{\tmop}[1]{\ensuremath{\operatorname{#1}}}

\begin{document}

\title{ Emergence and disruption of spin-charge separation in one-dimensional repulsive  fermions }

\author{Feng He}
\affiliation{State Key Laboratory of Magnetic Resonance and Atomic and Molecular Physics,
	Wuhan Institute of Physics and Mathematics,  APM, Chinese Academy of Sciences, Wuhan 430071, China}
\affiliation{University of Chinese Academy of Sciences, Beijing 100049, China.}

\author{Yu-Zhu Jiang}
\affiliation{State Key Laboratory of Magnetic Resonance and Atomic and Molecular Physics,
	Wuhan Institute of Physics and Mathematics, APM, Chinese Academy of Sciences, Wuhan 430071, China}

\author{Hai-Qing  Lin}
\email[]{haiqing0@csrc.ac.cn}
\affiliation{Beijing Computational Science Research Center, Beijing 100193, China
}
\affiliation{Department of Physics, Beijing Normal University, Beijing, 100875, China}

\author{Randall G. Hulet}
\affiliation{Department of Physics and Astronomy, and Rice Center for Quantum Materials,
Rice University, Houston, Texas 77251-1892, USA}

\author{Han Pu}
\affiliation{Department of Physics and Astronomy, and Rice Center for Quantum Materials,
Rice University, Houston, Texas 77251-1892, USA}

\author{Xi-Wen Guan}
\email[]{xiwen.guan@anu.edu.au}
\affiliation{State Key Laboratory of Magnetic Resonance and Atomic and Molecular Physics,
	Wuhan Institute of Physics and Mathematics, APM,  Chinese Academy of Sciences, Wuhan 430071, China}
\affiliation{Center for Cold Atom Physics, Chinese Academy of Sciences, Wuhan 430071, China}
\affiliation{Department of Theoretical Physics, Research School of Physics and Engineering,
	Australian National University, Canberra ACT 0200, Australia}


\date{\today}

%

\begin{abstract}
At low temperature,  collective excitations of one-dimensional (1D) interacting fermions  exhibit  spin-charge separation, a unique feature predicted by the Tomonaga-Luttinger liquid (TLL) theory, but a rigorous understanding remains challenging.
Using the thermodynamic Bethe Ansatz (TBA) formalism, we analytically derive universal properties of a 1D repulsive spin-1/2 Fermi gas with arbitrary interaction strength.  
We show how spin-charge separation emerges from the exact TBA formalism, and how it is disrupted by the interplay between the two degrees of freedom which brings us beyond the TLL paradigm. 
Based on the exact low-lying excitation spectra, we further evaluate the spin and charge dynamical structure factors (DSFs).  
The peaks of the DSFs exhibit distinguishable propagating velocities of spin and charge as functions of interaction strength, which can be observed by Bragg spectroscopy with ultracold atoms. 

\end{abstract}

\maketitle

Interacting quantum many-body systems with rich internal degrees of freedom usually pose a formidable challenge for  theoretical study. 
Understanding how interactions between fermions affects the state of a quantum  liquid at low temperatures has been an important topic for over fifty years, and many outstanding questions still remain. 
A wealth of approximate formalism has been developed to understand the universal low-energy physics. These include Landau's Fermi liquid theory \cite{Landau:2008,quantumliquid:2018}, density matrix renormalization group \cite{schollwock:2011density,kollath2005spin}, Green function approach \cite{green:book}, etc. 
In particular, the Tomonaga-Luttinger liquid (TLL) theory \cite{haldane:1981,Giamarchi:book,Imambekov:2012} describes the universal low-energy physics of strongly correlated systems  in one dimension (1D). 
The TLL usually refers to the collective motion of bosons that is significantly different from the free fermion nature  in the Fermi liquid. 

A hallmark of 1D physics is the splitting of low-lying excitations of interacting fermions into two separate TTLs, i.e., the separated quasiparticles carry either spin or charge. 
This phenomenon is known as  spin-charge separation. 
Usually, TTL physics can be directly obtained from the Bethe ansatz (BA) solutions \cite{Recati:PhysRevLett.90.020401,Guan:2012,mestyan2019spin,PhysRevB.101.035149}, where the particle-hole excitations have the same energy for a given momentum. 
This special feature of the TLL, however,  is disrupted  once backward scattering is included or when the system is strongly disturbed by thermal fluctuations at quantum criticality \cite{Sachdev_2001,Guan:RMP}.
Although the realizations of 1D cold atom  systems \cite{Kinoshita:2004,Kinoshita:2006,Paredes:2004,Haller:2009,Pagano:2014,liao:2010,Yang:2017,Hulet:2018,PhysRevLett.122.090601} have confirmed many  predictions from exactly solvable models, including recent studies on  the dynamical deconfinement of spin and charge on  1D lattices  \cite{vijayan2020time,hilker2017revealing,bohrdt2018angle,barfknecht2019dynamics},
an observation of the unique spin-charge separation still remains a long-standing challenge in experiments \cite{Kim:1996,auslaender2005spin}.
We naturally ask if spin-charge separation, its criticality, and behaviour beyond the TLL can be observed in ultracold atoms in a well controlled manner. 

In this letter, we aim to answer these questions and report on the universal properties of spin-charge separated and disrupted liquids in a repulsive spin-$1/2$ Fermi gas. 
We present analytical results of thermodynamic and magnetic properties of the system which essentially mark the spin-charge separated liquids below a lower critical temperature, the universal scaling behaviour of free fermion quantum criticality above an upper critical  temperature,  and  the disrupted quantum liquids in between. 
 We also evaluate exact  low-lying excitations which indicate the separation of particle-hole continuum in the charge  sector from the  two-spinon spectrum in the spin sector.  
Such separated spectra are exploited to calculate  the charge and spin dynamic structure factors (DSFs) and to probe the emergent phenomena such as spin-charge separation and fractional excitations in Fermi gases. 
%

{\em Yang-Gaudin model ---} The Hamiltonian of the 1D $\delta$-function interacting Fermi gas, the so-called Yang-Gaudin model \cite{Yang:1967,Gaudin:1967}, is given by 
\begin{eqnarray}
{\cal H}= -\sum_{i=1}^{N} \frac{\partial^2}{\partial x_i^2}+ 2c \sum_{\mathclap{1 \le i < j \leq N}} \delta (x_i - x_j) -HM -\mu N, \label{Ham}
\end{eqnarray}
where the total number of particles $N$ and the magnetization $M=(N_{\uparrow }-N_{\downarrow})/2$ are defined by the numbers of spin-up $N_{\uparrow}$ and spin-down $N_{\downarrow}$ fermions,
$H$ and $\mu$ denote the external magnetic field and the chemical potential, respectively. 
All quantities in (\ref{Ham}) are dimensionless where we have adopted a units system with $\hbar=2m=1$, here $m$ is the mass of the particle. We  also define the number density $n=N/L$ ($L$ being the length of the system).

In this paper we only consider the repulsive interaction with $c>0$.
The whole set of the exact BA wave functions, spectra and the associated BA equations were obtained by Yang in 1967 \cite{Yang:1967}.

The universal properties of the system can be derived from the thermodynamic Bethe ansatz (TBA) equations which, for the repulsive Fermi gas, are given by \cite{ Lai:1971,Lai:1973,Takahashi:1971}
\begin{eqnarray}
	\varepsilon (k) &=& k^2-\mu-\frac{H}{2}-T \sum_{n=1 }^{\infty}a_n*{\rm ln} [1+{\rm e}^{- \phi_n (\lambda)/T}], \label{wholeTBA1}\quad \\
	\phi_n (\lambda)&=& nH-T a_n*\ln [1+{\rm e}^{- \varepsilon (k)/T}]\nonumber \\
	&+& T\sum_{m=1}^{\infty} T_{mn}*{\rm ln} [1+{\rm e}^{- \phi_m (\lambda)/T}] \label{wholeTBA2}
\end{eqnarray}
where $*$ denotes the convolution, $\varepsilon(k)$ and $\phi_n(\lambda)$ are the dressed energies for the charge and the length-$n$ spin strings, respectively, with $k$'s and $\lambda$'s being the rapidities;
the integral kernel $a_n(k)=\frac{1}{2 \pi} \frac{n c}{ (nc)^2 /4+k^2}$, and the functions $T_{mn}$ are given in Refs.~\cite{Takahashi:1971,Guan:2012} (also see Supplemental Material \cite{Supp} for more detail). 
Once $\varepsilon(k)$ is obtained, we can calculate the pressure, i.e., the equation of state $p=\frac{T}{2 \pi}\int_{-\infty}^{\infty} {\rm ln} [1+{\rm e}^{- \varepsilon (k)/T}] {\rm d}k$, from which all other thermodynamic quantities of interest can be obtained~\cite{Supp}. 
The TBA equations  (\ref{wholeTBA1}) and (\ref{wholeTBA2}) reveal the full spin and thermal fluctuations controlled by the interplay between spin and charge.
%

%


\begin{figure}[t]
	\begin{center}	
	\includegraphics[width=1.0\linewidth]{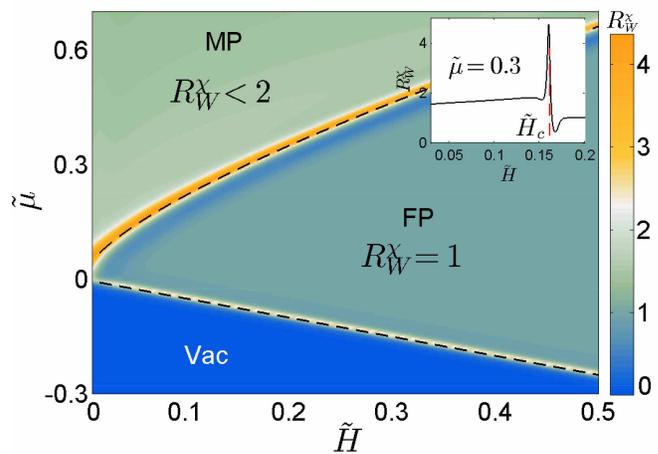}
	\end{center}
	\caption{ (color online) (a) Contour plot of Wilson Ratio (WR)  in $\tilde{\mu}-\tilde{H}$ plane for the  repulsive Fermi gas at  $\tilde{T}=0.005$. 
Here the dimensionless quantities $\tilde{T} =\frac{T}{|c|^2}, \, \tilde{\mu} = \frac{\mu}{|c|^2},  \, \tilde{H}=\frac{H}{|c|^2}$. 
The values of the WR given by Eq. (\ref{WR-TLL}) elegantly mark  three quantum phases: mixed phase (MP), full polarized phase (FP) and vacuum at zero temperature.  At low temperatures, the phase boundaries are indicated by sudden enhancements  of the WR,  which match well with the zero temperature phase boundaries (black dashed lines). The inset shows the WR vs magnetic field $\tilde{H}$ at $\tilde{\mu}=0.3$ and $\tilde{T}=0.005$, where a sudden enhancement of the WR is observed. }
	\label{fig1:phase}
\end{figure}

{\em Phase diagram and spin-charge separation ---}
Based on the configurations of spin orientations, the ground state phase diagram of a 1D repulsive Fermi gas in the $\tilde{\mu}$-$\tilde{H}$ plane contains three phases: vacuum, a mixed phase (MP) and a fully-polarized (FP) phase. 
The Wilson ratio (WR), defined as $R_W^{\chi} =\frac{4 }{3} \left( \frac{\pi k_B}{g \mu_B}\right)^2 \frac{\chi}{c_V/T}$, where $\chi$ is the magnetic susceptibility and $c_V$ the specific heat, captures the essence of the quantum liquid \cite{Wilson_1975,Guan:RMP,Guan:2013PRL}.
This ratio becomes temperature-independent in the TLL regime, while it displays a universal scaling behaviour in the vicinity of the quantum critical point, signalling a breakdown of the TLL. 
We show that the WR elegantly marks the low-temperature phase diagram, as can be seen in Fig.~\ref{fig1:phase}, and characterizes the TLL of spinons via the following relation \cite{Long-Paper}
\begin{equation}
R_W^{\chi} = \frac{2 v_c}{v_s+v_c}K_{s}.\label{WR-TLL}
\end{equation}
Here the Luttinger parameter $K_s =1$ at critical point and   $K_s <1$   in the MP phase. 
$R_W^{\chi}= 1$ for the FP phase. 
For the MP phase, we have  $ R_W^{\chi} <2$,  where the spin and the charge degrees of freedom dissolve into two separate TLLs with different speeds of propagation $v_s$ and $v_c$, respectively.

The spin-charge separation phenomenon for the Fermi gas describes a splitting of low-energy excitations in the spin and the charge sectors.
 %
Due to the limited capabilities to control interaction, spin density and temperature, unambiguously identifying  the spin-charge separation is extremely challenging. 
 %
 %
 Next, we derive rigorous results of spin-charge separation by means of the TBA equations (\ref{wholeTBA1}) and (\ref{wholeTBA2}) near and far from the quantum critical point (QCP) that separates the MP and the FP phases.


Throughout the MP phase with $H< H_c$, where $H_c$ is the critical field for a fixed chemical potential  (Fig.~\ref{fig1:phase}), 
we rigorously show \cite{Supp} that the pressure can, in general, be   given by
\begin{equation}
p-p_0=\frac{\pi T^2}{6} \left(  \frac{1}{v_c}+\frac{1}{v_s} \right), 
\end{equation}
where $p_0=\int_{-k_0}^{k_0}\varepsilon (k) dk$ is the pressure at $T=0$ and  the charge and spin velocities are given by 
\begin{equation} 
\label{vcvsdefination}
v_c=\frac{t_c}{2 \pi \rho_c(k_0)}\,,\;\;\;\; v_s=\frac{t_s}{2 \pi \rho_s(\lambda_0)}\,,
\end{equation}
respectively,  with $\rho_{c,s}$ being the distribution functions  at the Fermi points $k_0$ and $\lambda_0$  for the charge and the spin sector, (i.e., the points at which the dressed energies vanish), respectively; and $t_c$ and $t_s$ are the respective linear slopes of the dispersion at the Fermi points.  
We show that  $v_c$ and $v_s$ vary as functions of the external field $H$ for a fixed chemical potential. 
More detail is given in the Supplemental Material \cite{Supp}.
%
%
%
%
%
%
%
%

\begin{figure}[t]
	\begin{center}
	\includegraphics[width=1.0\linewidth]{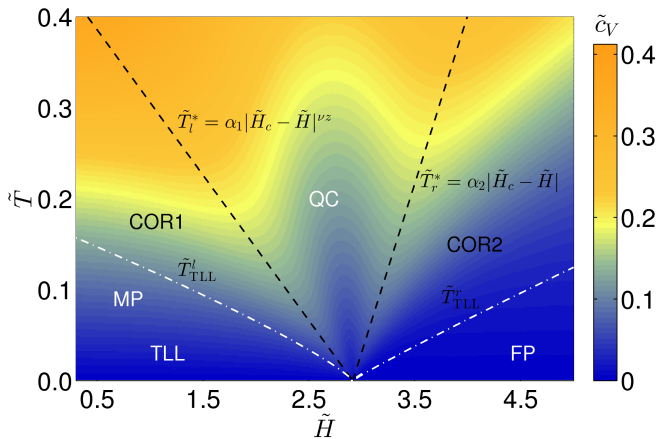}
	\end{center}
	\caption{(color online) Phase diagram in the $\tilde{T}$-$\tilde{H}$ plane: contour plot of specific heat. We set the  dimensionless chemical potentials $\tilde{\mu}=2.5$, $\tilde{H}_c=2.9145$. The black dashed lines denote the peak positions of specific heat, and the dot-dashed line shows the boundary of the linear $T$ dependence of specific heat. The crossover regions between QC and the TLL are labelled as COR1 and COR2.  }
	\label{Fig:phasediagram}
\end{figure}

{\em Quantum criticality and disrupted liquids ---} Understanding quantum criticality  and the disrupted Luttinger liquid provide a rich paradigm for many-body physics. 
In contrast to the spinless Bose gases \cite{Yang:2017},  
%
 the interplay between the spin and the charge degrees of freedom dramatically alters the critical behaviour of the system.
 For $c\to \infty$, the states of the system are highly degenerate and the spin sector becomes an incoherent free spin chain that does not exhibit magnetic ordering \cite{Takahashi_2005}.  
Here we consider a system with arbitrary interaction strength to obtain the universality class of quantum criticality encoding the interplay between spin and charge. 
Using the TBA equations (\ref{wholeTBA1}) and (\ref{wholeTBA2}), we find that  the phase transition occurs in the spin sector across the phase boundary between MP and FP phases, see \cite{Supp}. 
At finite temperatures, a quantum critical region (QC) fans out from the critical point, forming a critical cone in the $\tilde{T}$-$\tilde{H}$ plane, see Fig.~\ref{Fig:phasediagram}.
In the QC region, all thermodynamic quantities can be cast into universal scaling forms. Through an expansion of the length-1 spin string dressed energy equation (\ref{wholeTBA1}) and (\ref{wholeTBA2})  with an arbitrary interaction strength  at low temperatures, we  obtain the universal scaling function for the equation of states (pressure)  \cite{Supp}
\begin{eqnarray}
p-p_0&=&
\left\{
\begin{array}{lr}
-gT^{3/2} \tmop{Li}_{\frac{3}{2}} \left( -{\rm e}^{\frac{s_0 \Delta H }{T}} \right),\quad \text{for}\quad \mu=\mu_c, \\
-gT^{3/2} \tmop{Li}_{\frac{3}{2}} \left( -{\rm e}^{\frac{r_0 \Delta \mu}{T}} \right),\quad \text{for}\quad H=H_c,
\end{array}
\right.
\label{QCpressure}
 \end{eqnarray}
where $\Delta H =H_c-H$, $\Delta \mu=\mu_c-\mu$, $g= \frac{\arctan \left(2 k_0/c \right) }{ \pi^{3/2} \sqrt{ a }} $,  $s_0 =  1- \frac{1}{\pi} \arctan \left( \frac{2}{c} k_0 \right)$, $r_0 =-\frac{2}{\pi} \arctan \left( \frac{2}{c} k_0  \right) $ and $a$ is a constant determined by the critical chemical potential $\mu_c$ and the critical magnetic field $H_c$.
Here the Fermi momentum $k_0=\sqrt{\mu_c+H_c/2}$ is obtained from  the charge dressed energy condition $\varepsilon (k_0)=0$. 
The background pressure 
 \begin{eqnarray}
p_0&=&
\left\{
\begin{array}{lr}
\frac{\pi T^2}{6 \sqrt{\mu_c+H/2}}+ \frac{2}{3 \pi} \left( \mu_c +H/2 \right)^{3/2},\,\, \text{for}\quad \mu=\mu_c, \\
\frac{\pi T^2}{6 \sqrt{\mu+H_c/2}}+ \frac{2}{3 \pi} \left( \mu +H_c/2 \right)^{3/2} ,\,\, \text{for}\quad H=H_c,
\end{array}
\right.\label{pressure-BG-0}
 \end{eqnarray}
 reflects the regular part at quantum criticality. 
The correlation and dynamic critical exponents $\nu=1/2$ and $z=2$ are respectively read off by comparing  Eq.~(\ref{QCpressure}) with the universal scaling form $p-p_0=gT^{\frac{1}{z}+1} \mathcal{G}\left( \frac{s_0 \Delta H}{T^{1/\nu z}}, \frac{r_0 \Delta \mu}{T^{1/\nu z}} \right)$. 
These exponents also determine the two critical temperatures of the QC region  $T_{l}^{*} =\alpha_1 |H-H_c|^{\nu z}$ and $T_{r}^{*} =\alpha_2 |H-H_c|$, indicated by the two black dashed lines in Fig.~\ref{Fig:phasediagram}. Here $\alpha_{1,2}=s_0/y_{1,2}$ with $y_1=-1.5629$, $y_2=3.6205 $ are constants~~\cite{Long-Paper}.  %
Building on the exact scaling form of the pressure (\ref{QCpressure}), scaling functions of other thermodynamic quantities, such as magnetization, susceptibility, density, compressibility, and specific heat, can be evaluated in a straightforward way using standard statistical relations. 

Our result Eq.~(\ref{QCpressure}) provides not only a precise understanding of the emergent  criticality of spinons interplaying with charge~\cite{Supp}, 
  but also insightful perspectives of disrupted liquids beyond TLL. The interplay between the spin and the charge degrees of freedom leads to large deviations from the linear dispersion in both the spin and the charge sectors and to the disruption of the TLL in the crossover region $E_{\rm spin}\ll k_BT \ll E_{F}$, labelled as COR1 and COR2 in Fig.~\ref{Fig:phasediagram}. Here $E_{\rm spin}$ and $E_F$ are the energy of spin sector and Fermi energy, respectively. 
The crossover region COR1 coincides with the so-called  incoherent Luttinger liquid \cite{Fiete:2007,Cheianov:2004}.  We observe from $p_0$ in Eq.~(\ref{pressure-BG-0}) that the TLL nature only remains in the charge sector, while the  dilute deconfined spinons  become free fermion-like.
  These CORs reveal a coexistence of  liquid and gas-like states, more details see ~\cite{Long-Paper}.

\begin{figure}[t]
		\begin{center}
			\includegraphics[width=1.0 \linewidth]{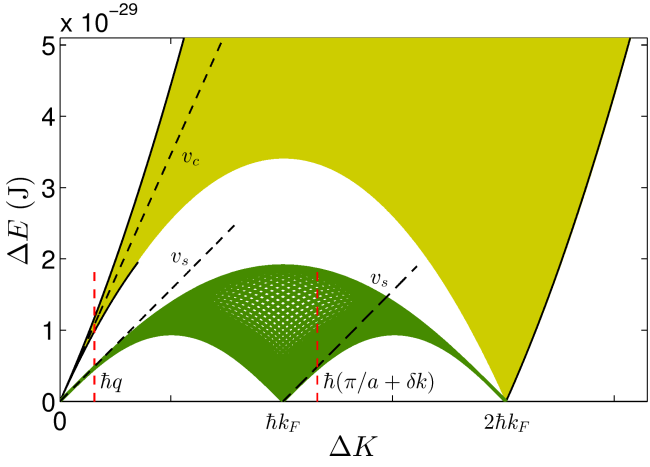}
		\end{center}  		
	\caption{{\small Exact low energy excitation spectra in charge (yellow green) and spin (dark green) at $\gamma =c/n= 5.03 \;(a_s=700a_0)$ with the Fermi surface $k_F=n\pi$, density $n=N/L=3 \times 10^6 \;(\rm 1/m)$, $\Delta E=\hbar \omega$. The yellow green   shows the particle-hole continuum excitation.  The black solid lines indicate  the thresholds of particle-hole excitation which remarkably manifest  the  free fermion-like  dispersion  (\ref{dispersion}) with  an  effective mass $m^* \approx 1.27 m$ at low energy. The black dished line in the charge excitation stands for the  charge velocity $v_c$. The dark green shows the two-spinon excitation, where the black dished  lines stand for  the spin velocities $v_s$ near $\Delta K=0$ and $\hbar k_F$, respectively. The two red dished  lines indicates  the positions of excitation  momenta in charge and spin sectors for Fig.~\ref{1Dhomspincharge}.  
	} }
	\label{spectrum}
\end{figure}

\begin{figure}[t]
		\begin{center}
			\includegraphics[width=1.0 \linewidth]{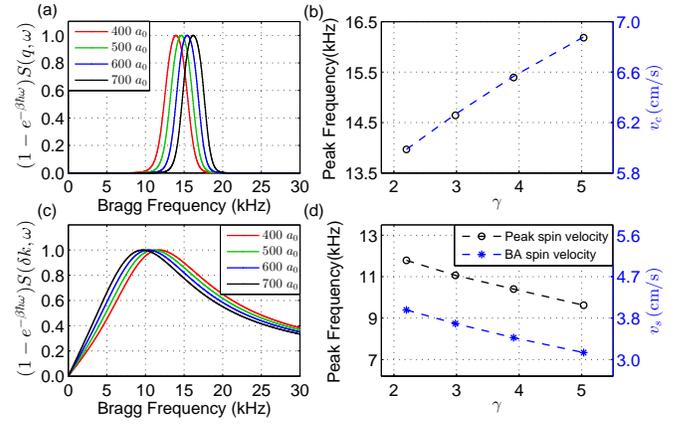}
		\end{center} 
\caption{(color online) Normalized charge and spin DSF's of a homogeneous Fermi gas with parameters corresponding to these of  \cite{Hulet:2018}:  length $L=20\;\mu m$, particle numbers $N=60$, temperature $T=120 \;{\rm nk}$, and various interaction strengths $a_s=400a_0$, \, $500a_0$,\, $600a_0$, \, $700a_0$. Here $a_s$ is the 3D scattering length, which is related to the 1D interaction strength by $c = -2 \hbar^2/m a_{\rm 1D}$ with $a_{1D}=\left( -a_{\perp }^{2}/2a_{s}\right) \left[ 1-C\left(
a_{s}/a_{\perp }\right) \right]$\cite{Oshanii_PRL_1998}. In converting to dimensional quantities, we have assumed the atoms are $^6$Li with transverse harmonic confinement $\omega_{\perp} =(2 \pi)198$ kHz.
(a) Normalized charge DSF [Eq.~(\ref{LDSorigin})] vs. Bragg frequency $\omega/2\pi$ at $q=1.47 \mu m^{-1}$. (b) The empty circles denote the peak frequency of each spectrum vs. $\gamma$. 
The corresponding peak charge velocity $\omega/q$ is given by the right axis. The dashed line is the charge sound velocity obtained from TBA.
 (c) Normalized spin DSF [Eq.~(\ref{TLLDSFSLL})] vs. Bragg frequency $\omega /2\pi$ at $\delta k=1.47 \mu m^{-1}$. (d) The empty circles denote the peak frequency of each spectrum vs. $\gamma$. 
 The corresponding peak spin velocity $\omega/\delta k$ is given by the right axis. Stars are spin sound velocity obtain from the TBA.
  }	\label{1Dhomspincharge}
	\end{figure}

{\em Exact low-lying excitations and dynamic structure factor ---}
%
 Solving the TBA equations (\ref{wholeTBA1}) and (\ref{wholeTBA2}), we obtain precisely the low-lying excitations in both spin and charge. As shown in Fig.~\ref{spectrum}, the excitations in the two sectors are separated from each other. 
The charge particle and hole excitations at low energy are given exactly  by
\begin{eqnarray}
\omega (q)&=&v_c |q| \pm \frac{\hbar q^2}{2 m^*}+\cdots \label{dispersion}
\end{eqnarray}
with $\frac{1}{2 m^*} =\frac{\varepsilon_c''(k_0)}{2(2 \pi \rho_c(k_0)^2}- \frac{\pi \rho_c'(k_0) \varepsilon_c'(k_0)}{(2 \pi \rho_c(k_0))^3}$, where $m^{*}$  is the effective mass, taking the form $m^* \approx m\left(1+ \frac{4\ln2}{\gamma}\right)$ as $\gamma \gg 1$ \cite{Supp}. 
For small $q$, the charge  excitation  can be well captured by the leading order in Eq.~(\ref{dispersion}), while the second term is irrelevant. 
%
The charge DSF in a 1D repulsive Fermi gas has been recently measured \cite{Hulet:2018,Yang:PhD} using the technique of Bragg spectroscopy \cite{Hoinka:2012,Brunello:2001}, where the key feature of free Fermi liquid was observed in the DSF and the speed of sound in the charge sector was measured. The charge DSF of a free homogeneous Fermi gas is already known to be \cite{cherny:2006polarizability}
\begin{equation}
\label{LDSorigin}
S(q,\omega)=\frac{{\rm Im} \chi (q, \omega,k_F,T,N)}{\pi (1-{\rm e}^{-\beta \hbar \omega})}.
\end{equation}
Based on the charge excitation spectrum (\ref{dispersion}), the interaction only modifies the effective mass with the Fermi point $k_F$ replaced by  $k_c=m^*v_c/\hbar$ \cite{Hulet:2018}.
 As a consequence, it will move the resonance position from $\omega=v_F q$ to $\omega\approx v_c q$ in the excitation spectrum.
Here we observe that for $T \rightarrow 0$, DSF $S(q,\omega)\ne 0$ only for  $\omega_{-} \leq \omega \leq \omega_{+}$, where $ \omega_{\pm}  =v_c |q| \pm \frac{\hbar q^2}{2 m^*}$ captures the dispersion (\ref{dispersion} ).  
Taking the setting for a gas of spin-balanced $^6\text{Li}$ with particle number $N=60$, several different values of interaction strength at temperature $T=120 \;{\rm nk}$,  tube length $L=20\;\mu m$, and $q=1.47 \mu m^{-1} \approx 0.15k_F$  \cite{Hulet:2018,Yang:PhD}, 
we demonstrate in Fig.~\ref{1Dhomspincharge}(a) the Bragg spectrum as a function of Bragg frequency. The peak frequency of the DSF signal is plotted in Fig.~\ref{1Dhomspincharge}(b) as a function of $\gamma$, from which we can read off the peak velocity defined as the ratio of peak frequency and $q$. As Fig.~\ref{1Dhomspincharge}(b) demonstrates, this peak velocity  is solely determined by  the charge sound velocity, whereas the effective mass affects the width of the DSF. 
Our results on charge velocity and its dependence of the interaction strength are consistent with the experimental measurement and analysis reported  in Ref.~\cite{Hulet:2018}. A more detailed study will be presented in near future~\cite{Long-Paper}. 
%
%

 In Fig.~\ref{spectrum}, we further show  that the low-lying excitation in the spin sector gives rise to  the  two-spinon excitation, which remarkably displays the low-energy behaviour of the Heisenberg spin-1/2 chain~\cite{Supp}.
 This two-spinon excitation spectrum holds for any finite interaction strength.
The spin DSF of the Fermi gas is associated with the spin-spin correlation described by an effective Heisenberg spin chain.
Near the Fermi momentum with wave number $\Delta K=\hbar (\pi/a+\delta k)$ with an effective  lattice constant $a=L/N$, the spin DSF is given by \cite{Schulz:1991,Giamarchi:book}
\begin{eqnarray}
S(\delta k, \omega)&=& \frac{1}{1-{\rm e}^{ -\beta \hbar \omega}}\frac{A_{LL}}{k_B T} {\rm Im} \left[ \rho \left( \frac{\hbar \omega+v_s \hbar \delta k}{4 \pi k_B T}\right) \right. \nonumber\\
&&\left. \times \rho \left( \frac{\hbar \omega-v_s \hbar \delta k}{4 \pi k_B T}\right) \right],\label{TLLDSFSLL}
\end{eqnarray}
where  $\rho(x)= \Gamma(1/4-ix)/\Gamma(3/4-ix)$, and $v_s$ is the spin velocity of the spin chain which can also be obtained from the second equation of (\ref{vcvsdefination}) in the strong interaction limit. Also, $A_{LL}=-c_{\perp}^2 \alpha /2$ is a constant with the length scale parameter $\alpha$ and a constant factor $c_{\perp}$. 
With the same setting for the above charge DSF, we show in Fig.~\ref{1Dhomspincharge}(c) and (d) the spin DSF signal and the spin peak velocity read off from its peak positions. As Fig.~\ref{1Dhomspincharge}(d) shows, unlike in the charge case, here the peak velocity does not coincide  with the spin sound velocity due to the peculiar  feature of the two-spinon excitation near  $\Delta K=\hbar \pi/a$ \cite{Supp}. However, both the spin peak and the sound velocities are almost linearly decreasing functions of $\gamma$, in contrast to the charge velocity dependence on $\gamma$. 
This is a clear and unambiguous demonstration of the spin-charge separation.  
The fractional excitations beyond the two-spinon DSF (\ref{TLLDSFSLL}) involve length-$n$ spin strings  (high order spinon process) in the spin imbalanced Fermi gas, see the TBA (\ref{wholeTBA1}) and (\ref{wholeTBA2}). 

{\em Summary ---} We have presented universal properties of the spin-charge separation  and disrupted liquids at and off quantum criticality. 
The  emergent liquid and gas-like  quantum phases  near QCP show a subtle  interplay between the spin and charge degrees of freedom.
The universal scaling functions, the crossover temperatures, as well as the DSFs deeply reveal the essence of the separated TLLs and their disruption which takes us beyond the spin-charge separation paradigm. 
Our method suggests a promising way to control fractional spin excitations, TLLs and magnetism   in ultracold atomic  systems with higher symmetries. 

\noindent

\section*{Acknowledgement}
The authors  thank Y. Y. Chen,  S. Cheng, A. del Campo and T. Giamarchi for helpful discussions. 
This work is supported by   the key NSFC grant No.\ 11534014 No.\ 11874393 and No.\ 1167420, and  the National Key R\&D Program of China  No.\ 2017YFA0304500.
HQL  acknowledges financial support from NSAF U1930402 and NSFC 11734002, as well as computational resources from the Beijing Computational Science Research Center. HP acknowledges supports from the US NSF and the Welch Foundation (Grant No. C-1669). 
RGH acknowledges support from an ARO MURI (Grant No. W911NF-14-1-0003), the US NSF (Grant No. PHY- 1707992), and the Welch Foundation (Grant No. C-1133).


\clearpage\newpage
\setcounter{figure}{0}
\setcounter{table}{0}
\setcounter{equation}{0}
\def\thefigure{S\arabic{figure}}
\def\thetable{S\arabic{table}}
\def\theequation{S\arabic{equation}}
\setcounter{page}{1}
\pagestyle{plain}

\begin{widetext}

\section*{Supplementary material: Spin-charge separated and disrupted liquids: Universal properties}

\begin{center}
\noindent{Feng He, Yu-Zhu Jiang, Hai-Qing Lin, Han Pu, Thierry Giamarchi,  Randy Hulet,  Xiwen Guan }
\end{center}

\section{ Yang-Gaudin model and Bethe Ansatz equations}
The Hamiltonian of the 1D $\delta$-function interacting Fermi gas  reads 
\begin{eqnarray}
{\cal H}= -\frac{\hbar^2}{2 m} \sum_{i=1}^{N} \frac{\partial^2}{\partial x_i^2}+ 2c \sum_{1 \le i < j \leq N} \delta (x_i - x_j) -HM^z -\mu N, \label{Ham}
\end{eqnarray}
where $N=N_{\uparrow }+N_{\downarrow}$ is the total number of particles, $M^z=(N_{\uparrow }-N_{\downarrow})/2$ is the magnetization with $N_{\uparrow}$ spin-up fermions and $N_{\downarrow}$ spin-down fermions, $H$ is the external magnetic field and $\mu$ is the chemical potential. The system is confined in a region with length $L$ and periodic boundary condition is assumed. Here we consider the repulsive interaction, i.e.,  $c >0$.
In the above Hamiltonian, the  coupling constant $c = -2 \hbar^2/m a_{\rm 1D}$  is determined  by the  1D scattering length, given by $a_{1D}=\left( -a_{\perp }^{2}/2a_{s}\right) \left[ 1-C\left(
a_{s}/a_{\perp }\right) \right]$\cite{Oshanii_PRL_1998}. In the following analysis we take $\hbar =1$, $n=N/L=1$ and $2m =1$, which defines our dimensionless unit system. 

The Bethe ansatz equations (BAE) for the repulsive Fermi gas with the periodic boundary condition are given by \cite{Takahashi:1999} 
\begin{eqnarray}
&&{\rm e}^{i k_j L} =\prod_{\alpha =1}^{M} \frac{k_j -\lambda_{\alpha} +ic/2}{k_j -\lambda_{\alpha} -ic/2}, \quad j=1,2,\cdots,N, \label{BA1} \\
&&\prod_{j=1}^N  \frac{\lambda_{\alpha}-k_j +ic/2}{\lambda_{\alpha}-k_j -ic/2}=-\prod_{\beta=1}^{M}\frac{\lambda_{\alpha}-\lambda_{\beta}+ic}{\lambda_{\alpha}-\lambda_{\beta}-ic}\,, \quad \alpha=1,2,\cdots,M.\label{BA2}
\end{eqnarray}
For repulsive interactions, the BAE do not admit complex roots in the charge degree of freedom $k_j$, whereas in the spin sector,  the spin string state are given by 
\begin{eqnarray}
\label{string}
\lambda_{\alpha}^{n,j}=\lambda_{\alpha}^{n}+\frac{ic}{2}(n+1-2j), \quad j=1,2,\cdots,n,
\end{eqnarray}
which are called the length-$n$ spin strings. 
Using this string hypothesis and the Yang-Yang approach, Lai \cite{Lai:1971, Lai:1973} and Takahashi \cite{Takahashi:1971} derived the thermodynamic Bethe ansatz (TBA) equations, which will be used for the study of the thermodynamics of the model. 
The TBA equations for the 1D repulsive Fermi gas are given by 
\begin{eqnarray}
	\varepsilon (k) &= &k^2-\mu-\frac{H}{2}-T \sum_{n=1 }^{\infty}a_n*{\rm ln} [1+{\rm e}^{- \phi_n (k)/T}], \nonumber  \\
	\phi_n (\lambda)&=& nH-T a_n*\ln [1+{\rm e}^{- \varepsilon (\lambda)/T}]+T\sum_{m=1}^{\infty} T_{mn}*{\rm ln} [1+{\rm e}^{- \phi_m (\lambda)/T}],\label{wholeTBA}
\end{eqnarray}
where
\begin{equation}
\label{fermian}
a_n(k)=\frac{1}{2 \pi} \frac{n c}{ (nc)^2 /4+k^2}. \nonumber \\
\end{equation}
and 
\begin{eqnarray}
T_{mn}(\lambda)=
\left\{
\begin{aligned}
&a_{|n-m|}(\lambda)+2a_{|n-m|+2}(\lambda)+\cdots + 2a_{m+n -2}(\lambda)+a_{m+n}(\lambda) &\text{for}\; m\ne n\\
&2a_2 (\lambda) +2a_{4}(\lambda)+\cdots+2a_{2n-2}(\lambda)+a_{2n}(\lambda) &\text{for}\; m= n
\end{aligned}
\right..
\end{eqnarray}
The pressure is given by
\begin{equation}
p=\frac{T}{2 \pi}\int_{-\infty}^{\infty} {\rm ln} [1+{\rm e}^{- \varepsilon (k)/T}] {\rm d}k \,, 
\end{equation}
from which all the thermal and magnetic quantities can be derived according to the standard statistical relations.

At low temperatures, $T\ll E_F$, we can safely neglect the contributions from the high strings  and just retain the leading length-1 string in the TBA equations. Under such an approximation, the low temperature TBA equations become 
\begin{eqnarray}
\label{1stringTBA}
\varepsilon (k) &=& k^2-\mu-\frac{H}{2}-T a_1*{\rm ln} [1+{\rm e}^{- \phi_1 (\lambda)/T}], \\
\phi_1 (\lambda)&=& H-T a_1*\ln [1+{\rm e}^{- \varepsilon (k)/T}]+T a_2*\ln [1+{\rm e}^{- \phi_1 (\lambda)/T}].
\end{eqnarray}
When temperature $T \rightarrow 0$, the TBA equations further reduce to
\begin{eqnarray}
\varepsilon_c^0 (k) &=&k^2 -\mu-H/2 + \int_{-\lambda_0}^{\lambda_0} a_1 (k-\lambda)\phi_s^0 (\lambda) {\rm d} \lambda, \label{chargezero}\\
\phi_s^0 (\lambda) &=&H+\int_{-k_0}^{k_0} a_1(\lambda-k) \varepsilon_c^0 (k) {\rm d}k -\int_{-\lambda_0}^{\lambda_0} a_2 (\lambda-\lambda')\phi_s^0 (\lambda') {\rm d}\lambda'.\label{spinzero}
\end{eqnarray}
The pressure for zero temperature is given by 
\begin{equation}
p_0=-\frac{1}{2 \pi}\int_{-k_0}^{k_0} \varepsilon_c^0 (k) {\rm d}k,
\end{equation}
where $k_0$ and $\lambda_0$ are zero points of dressed energies $\varepsilon$ and $\phi$ in charge and spin sectors, respectively.


\section{ Additivity rule of spin-charge separation}

Here we will derive analytically the additivity rule of spin-charge separation, as manifested in Eq.~(5) of the main text.

At low temperatures, the length-1 string TBA equations can be rewritten as 
\begin{eqnarray}
\varepsilon (k) &=& \varepsilon_c^0 (k) +\eta(k), \\
\phi_1 (\lambda) &=& \phi_s^0 (\lambda) +\gamma (\lambda),
\end{eqnarray}
where $\eta (k)$ and $\gamma (\lambda)$ are small corrections to the zero temperature charge and spin dressed energies, respectively.
The exact expression of the correction $\eta (k)$  can be evaluated by rewriting charge dressed energy as
\begin{eqnarray}
\varepsilon (k) &=&k^2 -\mu -\frac{H}{2} -T \int_{-\infty}^{\infty} a_1 (k-\lambda) \ln(1+ {\rm e}^{-\frac{\phi_1 (\lambda)}{T}}) {\rm d} \lambda \nonumber\\
&=&k^2 -\mu -\frac{H}{2} -T \int_{-\infty}^{\infty} a_1 (k-\lambda) \ln(1+ {\rm e}^{-\frac{|\phi_1 (\lambda)|}{T}}) {\rm d} \lambda + \int_{-\lambda_0}^{\lambda_0} a_1 (k-\lambda) \phi_1 (\lambda) {\rm d} \lambda \nonumber\\
&=&k^2 -\mu -\frac{H}{2} -T \int_{-\infty}^{\infty} a_1 (k-\lambda) \ln(1+ {\rm e}^{-\frac{|\phi_1 (\lambda)|}{T}}) {\rm d} \lambda + \int_{-\lambda_0}^{\lambda_0} a_1 (k-\lambda) (\phi_s^0 (\lambda)+\gamma (\lambda)) {\rm d} \lambda \nonumber\\
&=&\varepsilon_c^0(k)-T \int_{-\infty}^{\infty} a_1 (k-\lambda) \ln(1+ {\rm e}^{-\frac{|\phi_1 (\lambda)|}{T}}) {\rm d} \lambda + \int_{-\lambda_0}^{\lambda_0} a_1 (k-\lambda)\gamma (\lambda) {\rm d} \lambda \nonumber\\
&=& \varepsilon_c^0 (k) +\eta(k) 
\end{eqnarray}
Therefore one gets a new equation 
\begin{eqnarray}
\label{chargeeta}
\eta(k)= -T \int_{-\infty}^{\infty} a_1 (k-\lambda) \ln(1+ {\rm e}^{-\frac{|\phi_1 (\lambda)|}{T}}) {\rm d} \lambda +\int_{-\lambda_0}^{\lambda_0} a_1 (k-\lambda)\gamma (\lambda) {\rm d} \lambda 
\end{eqnarray}

Similarly, we repeat the calculation in the spin dressed energy equation, namely, 

\begin{eqnarray}
\phi_1 (\lambda) &=&H-T \int_{-\infty}^{\infty} a_1 (\lambda-k) \ln (1+ {\rm e}^{-\frac{\varepsilon (k)}{T}}) {\rm d} k +T \int_{-\infty}^{\infty} a_2(\lambda- \lambda') \ln (1+{\rm e}^{-\frac{\phi_1 (\lambda')}{T} })  {\rm d} \lambda'  \nonumber \\
&=&H-T \int_{-\infty}^{\infty} a_1 (\lambda-k) \ln (1+ {\rm e}^{-\frac{|\varepsilon (k)|}{T}}) {\rm d} k +\int_{-k_0}^{k_0} a_1 (k-\lambda) \varepsilon (k) {\rm d} k \nonumber \\
& &+T \int_{-\infty}^{\infty} a_2(\lambda- \lambda') \ln (1+{\rm e}^{-\frac{|\phi_1 (\lambda')|}{T} })  {\rm d} \lambda'-  \int_{-\lambda_0}^{\lambda_0} a_2(\lambda- \lambda') \phi_1 (\lambda')  {\rm d} \lambda'  \nonumber \\
&=&H-T \int_{-\infty}^{\infty} a_1 (\lambda-k) \ln (1+ {\rm e}^{-\frac{|\varepsilon (k)|}{T}}) {\rm d} k + \int_{-k_0}^{k_0} a_1 (k-\lambda) (\varepsilon_c^0 (k)+\eta(k)) {\rm d} k \nonumber \\
&& +T \int_{-\infty}^{\infty} a_2(\lambda- \lambda') \ln (1+{\rm e}^{-\frac{|\phi_1 (\lambda')|}{T} })  {\rm d} \lambda'-  \int_{-\lambda_0}^{\lambda_0} a_2(\lambda- \lambda') (\phi_s^0 (\lambda')+\gamma (\lambda'))  {\rm d} \lambda'  \nonumber \\
&=&\phi_s^0(\lambda) - T \int_{-\infty}^{\infty} a_1 (\lambda-k) \ln (1+ {\rm e}^{-\frac{|\varepsilon (k)|}{T}}) {\rm d} k +T \int_{-\infty}^{\infty} a_2(\lambda- \lambda') \ln (1+{\rm e}^{-\frac{|\phi_1 (\lambda')|}{T} })  {\rm d}  \lambda' \nonumber \\
&& + \int_{-k_0}^{k_0} a_1 (k-\lambda) \eta(k) {\rm d} k -  \int_{-\lambda_0}^{\lambda_0} a_2(\lambda- \lambda') \gamma (\lambda') {\rm d} \lambda'  \nonumber \\
&=&\phi_s^0 (\lambda) +\gamma (\lambda).
\end{eqnarray}
Thus we have 
\begin{eqnarray}
\label{spingamma}
\gamma (\lambda) &=&- T \int_{-\infty}^{\infty} a_1 (\lambda-k) \ln (1+ {\rm e}^{-\frac{|\varepsilon (k)|}{T}}) {\rm d} k +T \int_{-\infty}^{\infty} a_2(\lambda- \lambda') \ln (1+{\rm e}^{-\frac{|\phi_1 (\lambda')|}{T} })  {\rm d}  \lambda' \nonumber \\
&&+\int_{-k_0}^{k_0} a_1 (k-\lambda) \eta(k) {\rm d} k - \int_{-\lambda_0}^{\lambda_0} a_2(\lambda- \lambda') \gamma (\lambda') {\rm d} \lambda'.  \nonumber \\
\end{eqnarray}
The charge and spin dressed energies can be expanded at the Fermi points $k_0$ and $\lambda_0$
\begin{eqnarray}
\varepsilon (k) &=t_c (k-k_0) ,\quad t_c=\frac{{\rm d}\varepsilon(k) }{{\rm d} k} \bigg|_{k=k_0},  \nonumber \\
\phi_1 (\lambda) &= t_s (\lambda-\lambda_0), \quad t_s=\frac{{\rm d}\phi_1(\lambda) }{{\rm d} \lambda} \bigg|_{\lambda=\lambda_0}, 
\end{eqnarray}
where only the linear terms in the expansion are retained.

To expand the charge and spin dressed energies (\ref{chargeeta}) and (\ref{spingamma}) at the critical points, one can directly obtain
\begin{eqnarray}
\eta (k) &=&-\frac{\pi^2 T^2}{6 t_s} \left[ a_1(k-\lambda_0)+a_1 (k+\lambda_0)  \right] +\int_{-\lambda_0}^{\lambda_0} a_1(k-\lambda) \gamma(\lambda) {\rm d} \lambda, \\
\gamma (\lambda)&=&-\frac{\pi^2 T^2}{6 t_c} \left[ a_1(\lambda-k_0)+a_1 (\lambda+k_0)  \right] + \frac{\pi^2 T^2}{6 t_c} \left[ a_1(\lambda-\lambda_0)+a_1 (\lambda+\lambda_0)  \right] \nonumber \\
&& + \int_{-k_0}^{k_0} a_1 (k-\lambda) \eta(k) {\rm d} k -  \int_{-\lambda_0}^{\lambda_0} a_2(\lambda- \lambda') \gamma (\lambda') {\rm d} \lambda'
\end{eqnarray}
which can also be written as 
\begin{eqnarray}
\eta (k) &=&\eta^0 (k) +\int_{-\lambda_0}^{\lambda_0} a_1(k-\lambda) \gamma(\lambda) {\rm d} \lambda, \label{etazero}\\
\gamma (\lambda)&=&\gamma^0 (\lambda)+ \int_{-k_0}^{k_0} a_1 (k-\lambda) \eta(k) {\rm d} k -  \int_{-\lambda_0}^{\lambda_0} a_2(\lambda- \lambda') \gamma (\lambda') {\rm d} \lambda' ,\label{gammazero}
\end{eqnarray}
where we defined  $\eta_0$ and $\gamma_0$ as
\begin{eqnarray}
\label{etagamma}
\eta^0 (k) &=& -\frac{\pi^2 T^2}{6 t_s} \left[ a_1(k-\lambda_0)+a_1 (k+\lambda_0)  \right],  \nonumber\\
\gamma^0 (\lambda) &=& -\frac{\pi^2 T^2}{6 t_c} \left[ a_1(\lambda-k_0)+a_1 (\lambda+k_0)  \right] +\frac{\pi^2 T^2}{6 t_s} \left[ a_1(\lambda-\lambda_0)+a_1 (\lambda+\lambda_0)  \right].
\end{eqnarray}

Under a similar approximation, the pressure reduces to the following form 
\begin{eqnarray}
p&=&\frac{T}{2 \pi} \int_{-\infty}^{\infty} \ln [1+{\rm e}^{-\frac{\varepsilon (k)}{T}}] {\rm d} k =\frac{T}{2 \pi} \int_{-\infty}^{\infty} \ln [1+{\rm e}^{-\frac{|\varepsilon (k)|}{T}}] {\rm d} k-\frac{1}{2 \pi} \int_{-k_0}^{k_0} \varepsilon (k)  {\rm d} k \nonumber \\
&=&\frac{T}{2 \pi} \int_{-\infty}^{\infty} \ln [1+{\rm e}^{-\frac{|\varepsilon (k)|}{T}}] {\rm d} k-\frac{1}{2 \pi} \int_{-k_0}^{k_0} (\varepsilon_c^0 (k) +\eta (k))  {\rm d} k \nonumber\\
&=&p_0 +\frac{T}{2 \pi} \int_{-\infty}^{\infty} \ln [1+{\rm e}^{-\frac{|\varepsilon (k)|}{T}}] {\rm d} k-\frac{1}{2 \pi} \int_{-k_0}^{k_0} \eta (k)  {\rm d} k \nonumber \\
&=&p_0 +\frac{\pi T^2}{6 t_c}-\frac{1}{2 \pi} \int_{-k_0}^{k_0} \eta (k)  {\rm d} k.
\end{eqnarray}
Using the spin and charge densities
\begin{eqnarray}
\rho_c (k) &=&\frac{1}{2 \pi}+ \int_{-\lambda_0}^{\lambda_0}a_1 (k-\lambda) \rho_s(\lambda) {\rm d}\lambda,  \label{zerodensitycharge}\\
\rho_s (\lambda)&=&\int_{-k_0}^{k_0}  a_1 (\lambda-k) \rho_c (k) {\rm d}k - \int_{-\lambda_0}^{\lambda_0} a_2 (\lambda-\lambda')\rho_s (\lambda') {\rm d} \lambda', \label{zerodensityspin}
\end{eqnarray}
and the  expressions (\ref{etazero}) and (\ref{gammazero}), we then multiply  (\ref{etazero}) with (\ref{zerodensitycharge}) and integrate with $k$ 
\begin{eqnarray}
\label{midcalcu}
\frac{1}{2 \pi} \int_{-k_0}^{k_0} \eta (k) {\rm d}k+ \int_{-k_0}^{k_0} \int_{-\lambda_0}^{\lambda_0} a_1(k-\lambda) \eta(k)\rho_s(\lambda)  {\rm d}k  {\rm d}\lambda = \int_{-k_0}^{k_0} \eta^0 (k) \rho_c (k) {\rm d}k + \int_{-k_0}^{k_0} \int_{-\lambda_0}^{\lambda_0} a_1(k-\lambda) \gamma (\lambda) \rho_c(k) {\rm d}k  {\rm d}\lambda. 
\end{eqnarray}

Substituting  (\ref{gammazero}) and (\ref{zerodensityspin}) to the right hand side (r.h.s) of above equation (\ref{midcalcu}), then we have 
\begin{eqnarray}
\label{rightside}
\text{r.h.s}&= & \int_{-k_0}^{k_0} \eta^0 (k) \rho_c (k) {\rm d}k + \int_{-\lambda_0}^{\lambda_0}\left[ \rho_s(\lambda)+ \int_{-\lambda_0}^{\lambda_0} a_2 (\lambda-\lambda')\rho_s (\lambda') {\rm d} \lambda' \right] \gamma (\lambda) {\rm d}\lambda \nonumber \\
&= &\int_{-k_0}^{k_0} \eta^0 (k) \rho_c (k) {\rm d}k +  \int_{-\lambda_0}^{\lambda_0}  \rho_s (\lambda) \gamma^0 (\lambda)  {\rm d}\lambda + \int_{-\lambda_0}^{\lambda_0}\int_{-k_0}^{k_0} a_1 (k-\lambda) \eta(k) \rho_s (\lambda) {\rm d}\lambda  {\rm d} k \nonumber \\
&& - \int_{-\lambda_0}^{\lambda_0} \int_{-\lambda_0}^{\lambda_0} a_2(\lambda- \lambda') \gamma (\lambda') \rho_s (\lambda) {\rm d}\lambda  {\rm d} \lambda' + \int_{-\lambda_0}^{\lambda_0} \int_{-\lambda_0}^{\lambda_0} a_2(\lambda-\lambda') \gamma (\lambda) \rho_s (\lambda') {\rm d}\lambda  {\rm d} \lambda'.
\end{eqnarray}
The comparison between the left hand side of (\ref{midcalcu}) and (\ref{rightside})  gives the following relation 
\begin{eqnarray}
\label{corssingrelation}
\frac{1}{2 \pi}\int_{-k_0}^{k_0} \eta (k)  {\rm d} k &=&\int_{-k_0}^{k_0} \eta^0 (k) \rho_c (k) {\rm d} k +\int_{-\lambda_0}^{\lambda_0} \gamma^0 (\lambda) \rho_s (\lambda) {\rm d} \lambda.
\end{eqnarray}
Using the explicit expression (\ref{etagamma}), we obtain
\begin{eqnarray}
\frac{1}{2 \pi}\int_{-k_0}^{k_0} \eta (k)  {\rm d} k 
&=&-\frac{\pi^2 T^2}{6 t_s} \int_{-k_0}^{k_0} \left[ a_1(k-\lambda_0)+a_1 (k+\lambda_0)  \right] \rho_c (k) {\rm d} k   -\frac{\pi^2 T^2}{6 t_c} \int_{-\lambda_0}^{\lambda_0}\left[ a_1(\lambda-k_0)+a_1 (\lambda+k_0)  \right] \rho_s (\lambda) {\rm d} \lambda \nonumber \\
&&+\frac{\pi^2 T^2}{6 t_s} \int_{-\lambda_0}^{\lambda_0}\left[ a_1(\lambda-\lambda_0)+a_1 (\lambda+\lambda_0)  \right] \rho_s(\lambda) {\rm d} \lambda \nonumber \\
&=&-\frac{\pi^2 T^2}{6 t_s}(2\rho_s (\lambda_0))-\frac{\pi^2 T^2}{6 t_c}(2\rho_c (k_0)-\frac{1}{\pi}) \nonumber \\
&=&-\frac{\pi^2 T^2}{3 t_s}\rho_s (\lambda_0)-\frac{\pi^2 T^2}{3 t_c}\rho_c (k_0)+\frac{\pi T^2}{6 t_c}  \label{linearT}. 
\end{eqnarray}
In the derivation above, the symmetric property of density equations are used. To see this clearly, we show the density symmetry relations below.

 For the charge and spin densities, at Fermi point $k_0$ and $\lambda_0$, we have
\begin{eqnarray}
\rho_c (k_0) &=&\frac{1}{2 \pi}+ \int_{-\lambda_0}^{\lambda_0}a_1 (k_0-\lambda) \rho_s(\lambda) {\rm d}\lambda \label{rhock0},\\
\rho_c (-k_0) &=&\frac{1}{2 \pi}+ \int_{-\lambda_0}^{\lambda_0}a_1 (-k_0-\lambda) \rho_s(\lambda) {\rm d}\lambda \label{rhock00}. \\
\rho_s (\lambda_0)&=&\int_{-k_0}^{k_0}  a_1 (\lambda_0-k) \rho_c (k) {\rm d}k - \int_{-\lambda_0}^{\lambda_0} a_2 (\lambda_0-\lambda')\rho_s (\lambda') {\rm d} \lambda', \label{rhosk0}\\
\rho_s (-\lambda_0)&=&\int_{-k_0}^{k_0}  a_1 (-\lambda_0-k) \rho_c (k) {\rm d}k - \int_{-\lambda_0}^{\lambda_0} a_2 (-\lambda_0-\lambda')\rho_s (\lambda') {\rm d} \lambda'. \label{rhosk00}
\end{eqnarray}
Moreover, (\ref{rhock00}) and (\ref{rhosk00}) can also be rewritten as
\begin{eqnarray}
\rho_c (k_0) &=&\frac{1}{2 \pi}+ \int_{-\lambda_0}^{\lambda_0}a_1 (k_0+\lambda) \rho_s(\lambda) {\rm d}\lambda  \label{rhock000} . \\
\rho_s (\lambda_0)&=&\int_{-k_0}^{k_0}  a_1 (\lambda_0+k) \rho_c (k) {\rm d}k - \int_{-\lambda_0}^{\lambda_0} a_2 (\lambda_0+\lambda')\rho_s (\lambda') {\rm d} \lambda'. \label{rhosk000}
\end{eqnarray}
since both charge and spin densities  are even functions of $k$ and $\lambda$ respectively.

Summing up equations (\ref{rhock0}) and (\ref{rhock000}), (\ref{rhosk0}) and (\ref{rhosk000}), we obtain 
\begin{eqnarray}
&&\int_{k_0}^{k_0} [a_1 (k_0-\lambda)+a_1 (k_0+\lambda)] \rho_s(\lambda) {\rm d}\lambda =2 \rho_c (k_0)-\frac{1}{\pi} \label{rhocplus}. \\
&&\int_{-k_0}^{k_0}  [a_1 (\lambda_0-k)  + a_1 (\lambda_0+k ) ] \rho_c (k) {\rm d} k- \int_{-\lambda_0}^{\lambda_0} [a_2 (\lambda_0-\lambda')+a_2 (\lambda_0+\lambda')]\rho_s  (\lambda') {\rm d} \lambda =2\rho_s (\lambda). \label{rhosplus}
\end{eqnarray}

According to (\ref{rhocplus}),  (\ref{rhosplus}) and the relation (\ref{linearT}), the pressure is given by 
\begin{eqnarray}
p-p_0&=&\frac{\pi T^2}{6 t_c}-\frac{1}{2 \pi} \int_{-k_0}^{k_0} \eta (k)  {\rm d} k \nonumber \\
&=&\frac{\pi T^2}{6 t_c}+\frac{\pi^2 T^2}{3 t_s}\rho_s (\lambda_0)+\frac{\pi^2 T^2}{3 t_c}\rho_c (k_0)-\frac{\pi T^2}{6 t_c}  \nonumber \\
&=&\frac{\pi^2 T^2}{3 t_s}\rho_s (\lambda_0)+\frac{\pi^2 T^2}{3 t_c}\rho_c (k_0).
\end{eqnarray}
By definition, the charge and the spin velocities read \cite{Lee:2012}
\begin{equation}
\label{vcvsdefination}
v_c=\frac{t_c}{2 \pi \rho_c(k_0)}, \quad  v_s=\frac{t_s}{2 \pi \rho_s(\lambda_0)},
\end{equation}
such that the low temperature correction to the pressure is
\begin{equation}
p-p_0=\frac{\pi T^2}{6} \left(  \frac{1}{v_c}+\frac{1}{v_s} \right).
\end{equation}
which is Eq.~(5) in the main text.
This represents a rigorous proof of the additivity rule of the leading temperature contributions to the free energy (or pressure). These corrections reflect the characteristic linear dispersion in the spin and the charge degrees of freedom.  
The specific heat can be obtained readily as
\begin{eqnarray} 
\label{cVzerot}
c_{V}= \frac{\pi T}{3} \left(  \frac{1}{v_c}+\frac{1}{v_s} \right).
\end{eqnarray}
The expressions of the pressure and the specific heat show the universal low temperature thermodynamics in terms of two separated degrees of freedom: the spin and the charge.

\section{Scaling functions at quantum criticality}
In this section, we include more details on the derivation of Eq.~(7) in the main text, which is one of the key results of our work.

In fact, it is a formidable task to derive universal scaling functions for the phase transition from the MP phase to the FP phase in an analytical fashion. 
Based on the fact that quantum phase transition occurs at zero temperature,
universal scaling behaviour can be derived in the vicinity of the critical point at low temperature. 
From the TBA equations (\ref{wholeTBA}), we observe that near the critical point the length-1 spin string pattern dominates the TBA equations at low temperature. 
Near the critical point, the spin dressed energy $\phi(\lambda)$ only has  a small negative part, which mainly determines the charge and the spin dressed energies near the critical point at low temperatures. 
Therefore, we can expand  the integration  kernel $a_n(k-\lambda)$ in terms of the  functions of small variables $\lambda$.
This leads to a deconvolution in the whole TBA equations  (\ref{wholeTBA}).
Therefore we can calculate the scaling functions  by approximating the spin dressed energy in terms of the power  of $\lambda^n$.
The whole approximation procedure is rather complicated.
Here we prefer to present a few key steps for a demonstration of the validity of our scaling functions, more detailed study will  be presented elsewhere  \cite{Long-Paper}.

In order to obtain universal thermodynamics, we first expand  the kernel function 
\begin{eqnarray}
a_n(k-\lambda) = \frac{1}{2\pi}\frac{nc}{(nc)^2/4+(k-\lambda)^2} 
\approx \frac{nc}{2\pi} \frac{1}{(nc)^2/4+k^2} \left[ 1+\frac{2k\lambda-\lambda^2}{(nc)^2/4+k^2}+\frac{4k^2\lambda^2 -4k\lambda^3+ \lambda^4}{\left((nc)^2/4+k^2 \right)^2}+\cdots  \right]. \nonumber \\ 
\end{eqnarray}
Here near the critical point, the conditions $c, k \gg \lambda$ hold. 
Up to the order of  $\sim O(\lambda^2)$, the TBA equations are reduced to the form 
\begin{eqnarray}
\label{spindressedenergyexpansion}
\varepsilon(k) &=& k^2 -\mu -\frac{H}{2} -T \int_{-\infty}^{\infty} a_1 (k-\lambda) \ln(1+ {\rm e}^{-\frac{\phi_1 (\lambda)}{T}}) {\rm d} \lambda \nonumber\\
&=& k^2 -\mu -\frac{H}{2} -\frac{Tc}{2\pi} \frac{1}{c^2/4+k^2} \int_{-\infty}^{\infty} \ln(1+ {\rm e}^{-\frac{\phi_1 (\lambda)}{T}}) {\rm d} \lambda \nonumber \\
&&+ \frac{Tc}{2\pi} \left[ \frac{1}{(c^2/4+k^2)^2}- \frac{4k^2}{(c^2/4+k^2)^3}\right] \int_{-\infty}^{\infty} \lambda^2 \ln(1+ {\rm e}^{-\frac{\phi_1 (\lambda)}{T}}) {\rm d} \lambda,   \\
\phi_1(\lambda) &=&H-T \int_{-\infty}^{\infty} a_1 (\lambda-k) \ln (1+ {\rm e}^{-\frac{\varepsilon (k)}{T}}) {\rm d} k +T \int_{-\infty}^{\infty} a_2(\lambda- \lambda') \ln (1+{\rm e}^{\frac{\phi (\lambda')}{T} })  {\rm d} \lambda', \nonumber \\
&\approx& b+a\lambda^2+c_1+c_2 \lambda^2
\end{eqnarray}
where $b=H-b_1$, and we have defined the following   factors 
\begin{eqnarray}
b_1 &=&  \frac{Tc}{2 \pi} \int_{-\infty}^{\infty} \frac{1}{c^2/4+k^2} \ln (1+ {\rm e}^{-\frac{\varepsilon (k)}{T}}) {\rm d} k, \\
a&=&\frac{Tc}{2 \pi} \int_{-\infty}^{\infty} \left[\frac{1}{(c^2/4+k^2)^2}- \frac{4k^2}{(c^2/4+k^2)^3} \right] \ln (1+ {\rm e}^{-\frac{\varepsilon (k)}{T}}) {\rm d} k, \\
c_1 &= & \frac{Tc}{\pi} \int_{-\infty}^{\infty} \frac{1}{c^2+\lambda'^2} \ln (1+ {\rm e}^{-\frac{\phi (\lambda')}{T}}) {\rm d} \lambda',  \\
c_2 &=&-\frac{Tc}{\pi} \int_{-\infty}^{\infty} \left[\frac{1}{(c^2+\lambda'^2)^2}- \frac{4 \lambda'^2}{(c^2+\lambda'^2)^3} \right] \ln (1+ {\rm e}^{-\frac{\phi (\lambda')}{T}}) {\rm d} \lambda'. 
\end{eqnarray}
The integrations  in the functions $b_1$ and $a$ are very hard to calculate. 
Like the approximation made in the previous section, we  separate the negative and positive parts of charge dressed energy to approximate the integration in $b_1$ and $a$. 
This approximation turns out  to be very efficient  near a phase transition. 

To this end, we assume that $\varepsilon (k) = t_c (k-k_T)$ near critical point, where $t_c=\partial \varepsilon(k)/\partial k |_{k=k_T}$,  $k_T$ denote the Fermi point of the charge dressed energy at finite temperatures.  Then we get 
\begin{eqnarray}
b_1 &=&  \frac{Tc}{2 \pi} \int_{-\infty}^{\infty} \frac{1}{c^2/4+k^2} \ln (1+ {\rm e}^{-\frac{\varepsilon (k)}{T}}) {\rm d} k \nonumber \\
&=&\frac{c}{2 \pi}\left[ \frac{\pi^2 T^2}{3t_c} \frac{1}{c^2/4+k_T^2} -\int_{-k_T}^{k_T} \frac{\varepsilon(k)}{c^2/4+k^2} {\rm d} k \right].
\end{eqnarray}
The first term in $b_1$ is negligible  because at the low temperatures,  the quantities  $k_T$ and $t_c$ can be large along phase boundary. 
At quantum criticality, it is safe to work out the thermodynamics in the $T\to 0$ limit. Therefore for getting a close form of scaling function, we may take $\varepsilon (k)\approx k^2-\mu +H/2$ in $b_1$. It follows that 
\begin{eqnarray}
b_1 &\approx& -\frac{c}{2 \pi}\int_{-k_T}^{k_T} \frac{\varepsilon (k)}{c^2/4+k^2} {\rm d} k 
=\left(\frac{c^2}{2 \pi}+\frac{2 }{\pi}k_T^2 \right)\arctan(\frac{2}{c}k_T)-\frac{c}{\pi} k_T. 
\end{eqnarray}
The quantity  $k_T$ near critical point is given by 
\begin{eqnarray}
k_T =\sqrt{\mu+H/2} &=\sqrt{\mu_c+H_c/2} \left( 1-\frac{\Delta \mu}{\mu_c +H_c/2} -\frac{\Delta H /2}{\mu_c +H_c/2} \right)^{1/2}, 
\end{eqnarray}
where we have defined $\Delta \mu = \mu_c -\mu$ and $\Delta H=H_c-H$.  We further obtain 
\begin{eqnarray}
b_1 &\approx&  H_c -\frac{2}{\pi} \arctan ( \frac{2}{c} k_0 ) \Delta \mu-\frac{1}{\pi} \arctan ( \frac{2}{c} k_0 )\Delta H,\nonumber\\
b &=&H-b_1 \approx \left[ \frac{1}{\pi} \arctan ( \frac{2}{c} k_0 ) -1 \right] \Delta H+\frac{2}{\pi} \arctan ( \frac{2}{c} k_0 ) \Delta \mu \nonumber\\
&=&-s_0 \Delta H -r_0 \Delta \mu, 
\end{eqnarray}
where 
\begin{eqnarray}
s_0 =  1- \frac{1}{\pi} \arctan \left( \frac{2}{c} k_0 \right) \quad  r_0 =-\frac{2}{\pi} \arctan \left( \frac{2}{c} k_0  \right). 
\end{eqnarray}
Then we can calculate the function $\phi_1 (\lambda)$
\begin{equation}
\label{wholephi}
\phi_1 (\lambda)=b+a\lambda^2+c_1+c_2 \lambda^2 \approx -s_0\Delta H  -r_0 \Delta \mu+(a+c_2) \lambda^2+c_1
\end{equation}
with the constants 
\begin{eqnarray}
c_1 &=& \frac{Tc}{\pi} \int_{-\infty}^{\infty} \frac{1}{c^2+\lambda'^2} \ln (1+ {\rm e}^{-\frac{\phi (\lambda')}{T}}) {\rm d} \lambda',\\
c_2 &=&-\frac{Tc}{\pi} \int_{-\infty}^{\infty} \left[\frac{1}{(c^2+\lambda'^2)^2}- \frac{4 \lambda'^2 }{(c^2+\lambda'^2)^3} \right] \ln (1+ {\rm e}^{-\frac{\phi (\lambda')}{T}}) {\rm d} \lambda'.
\end{eqnarray}
From  Eq. (\ref{wholephi}), we can work out the constant 
\begin{eqnarray}
a&\approx& -\frac{c}{2 \pi}\int_{-k_T }^{k_T} \left[\frac{1}{(c^2/4+k^2)^2}- \frac{4k^2}{(c^2/4+k^2)^3} \right] \varepsilon (k)   {\rm d} k   
\end{eqnarray}
Without losing generality,  we fix the chemical potential $\mu =\mu_c$ in our  discussion.  The negative part of the spin dressed energy $\phi_1(\lambda)$ is very small in vicinity of the critical point. 
We observe that  $a\lambda^2+b$ from the charge degree of freedom is much larger than the spin fluctuation $ c_2 \lambda^2 +c_1$.
Thus we can treat $\phi_1(\lambda)=a\lambda^2 +b$ as the initial value in iteration.
 It follows that 
\begin{eqnarray}
\label{c1expansion}
c_1 &\approx&  -\frac{T^{3/2} }{\sqrt{\pi a} c} \tmop{Li}_{3/2}\left( -{\rm e}^{\frac{s_0 \Delta H}{T}} \right)  +\frac{ T^{5/2}}{2 \sqrt{\pi} a^{3/2} c^3} \tmop{Li}_{5/2}\left( -{\rm e}^{ \frac{s_0 \Delta H}{T} } \right), \\
c_2 &
\approx& \frac{T^{3/2}}{ \sqrt{\pi a } c^3} \tmop{Li}_{3/2}\left( -{\rm e}^{s_0 \Delta H/T} \right) - \frac{3 T^{5/2}}{ \sqrt{\pi}a^{3/2} c^3} \tmop{Li}_{5/2}\left( -{\rm e}^{s_0 \Delta H/T} \right). 
\end{eqnarray}
Using the function $\phi_1(\lambda)$, we further calculate  the  dressed energy of the charge
\begin{eqnarray}
\varepsilon(k) 
&\approx &k^2 -\mu_c -\frac{H}{2}+\frac{D_1}{c^2/4+k^2}-\left[ \frac{1}{(c^2/4+k^2)^2}- \frac{4k^2}{(c^2/4+k^2)^3}\right] D_2, 
\end{eqnarray}
where we denote 
\begin{eqnarray}
D_1 &=&\frac{T^{3/2} c }{2 \sqrt{\pi (a+c_2)}}  \tmop{Li}_{\frac{3}{2}} \left( -{\rm e}^{\frac{s_0 \Delta H -c_1}{T}} \right), \nonumber \\
D_2 &=&\frac{T^{5/2}c}{4 \sqrt{\pi}(a+c_2)^{3/2}} \tmop{Li}_{\frac{5}{2}}\left( -{\rm e}^{\frac{s_0\Delta H -c_1}{T}} \right). 
\end{eqnarray}
Here  $D_1$ and $D_2$ are very small and $D_1 \gg D_2$ at low temperatures. 
Finally, we obtain the pressure  Eq. (7) of the system near the phase transition from MP phase to FP phase in the main text 
\begin{eqnarray}
p &\approx& p_0-  \frac{ \arctan \left(\frac{2}{c} k_0 \right) T^{3/2} }{ \pi^{3/2} \sqrt{ (a+c_2)}}  \tmop{Li}_{\frac{3}{2}} \left( -{\rm e}^{\frac{s_0 \Delta H -c_1}{T}} \right) +\frac{ T^{5/2}}{4 \pi^{3/2} (a+c_2)^{3/2} } \frac{c k_0}{ \left( c^2/4+k_T^2 \right)^2}  \tmop{Li}_{\frac{5}{2}} \left( -{\rm e}^{\frac{s_0 \Delta H -c_1}{T}} \right), \label{pressure-QC} 
\end{eqnarray}
where the  pressure $p_0$ is given by 
\begin{eqnarray}
\label{backgoudprssure}
p_0 = \frac{\pi T^2}{6 t_c}+ \frac{2}{3 \pi} \left( \mu_c +\frac{H}{2} \right)^{3/2} =  \frac{\pi T^2}{6 v_c}+ \frac{2}{3 \pi} \left( \mu_c +\frac{H}{2} \right)^{3/2} = p_0^{\rm Liquid} +p_0^{\rm BG}.
\end{eqnarray}
In the above expression  $p_0^{\rm BG}=\frac{2}{3 \pi} \left( \mu_c +\frac{H}{2} \right)^{3/2}$ can be regarded as the background part of charge, whereas $  p_0^{\rm Liquid}=\frac{\pi T^2}{6 v_c}$ 
denotes the Luttinger liquid contribution from charge degrees of freedom. 
Whereas the Luttinger liquid in the spin sector dissolves into the free fermion criticality, i.e. 
the pressure (\ref{pressure-QC}) is given by  a universal scaling form of the  equation of states
\begin{eqnarray}
p=p_0^{\rm Liquid}+p_0^{\rm BG}+T^{\frac{1}{z}+1} \mathcal{G}\left( \frac{s_0 \Delta H}{T^{1/\nu z}} \right). 
\end{eqnarray} 
Consequently, the scaling functions of  all  thermodynamic quantities can be derived based on this exact expression of the equation of states. We will present a more detailed study of various scaling functions in Ref.~\cite{Long-Paper}.

\section{Excitation spectra and  dynamic structure factors}

\subsubsection{Charge dynamic structure factor}
The dynamic structure factor (DSF) for the 1D  repulsive Fermi gas (\ref{Ham})  has not been analytically  studied  yet. The charge DSF of 1D non-interacting homogeneous free Fermi gas is given by \cite{Cherny:2006}
\begin{eqnarray}
\label{LDSorigin}
S(q,\omega)=\frac{{\rm Im} \chi (q, \omega,k_F,T,N)}{\pi (1-{\rm e}^{-\beta \hbar \omega})},
\end{eqnarray}
where the dynamic polarizability is given by 
\begin{eqnarray}
\chi(q,\omega,k_F,T,N)=\sum_k \frac{n_{k+q/2}-n_{k-q/2}}{\hbar \omega -\hbar^2 k q/m^*+i\eta}.
\end{eqnarray} 
In the above equations, $m^*$ denotes the effective mass of quasiparticles. 

\begin{figure}[tbp]
	\centering
	{\includegraphics[width=3.4in]{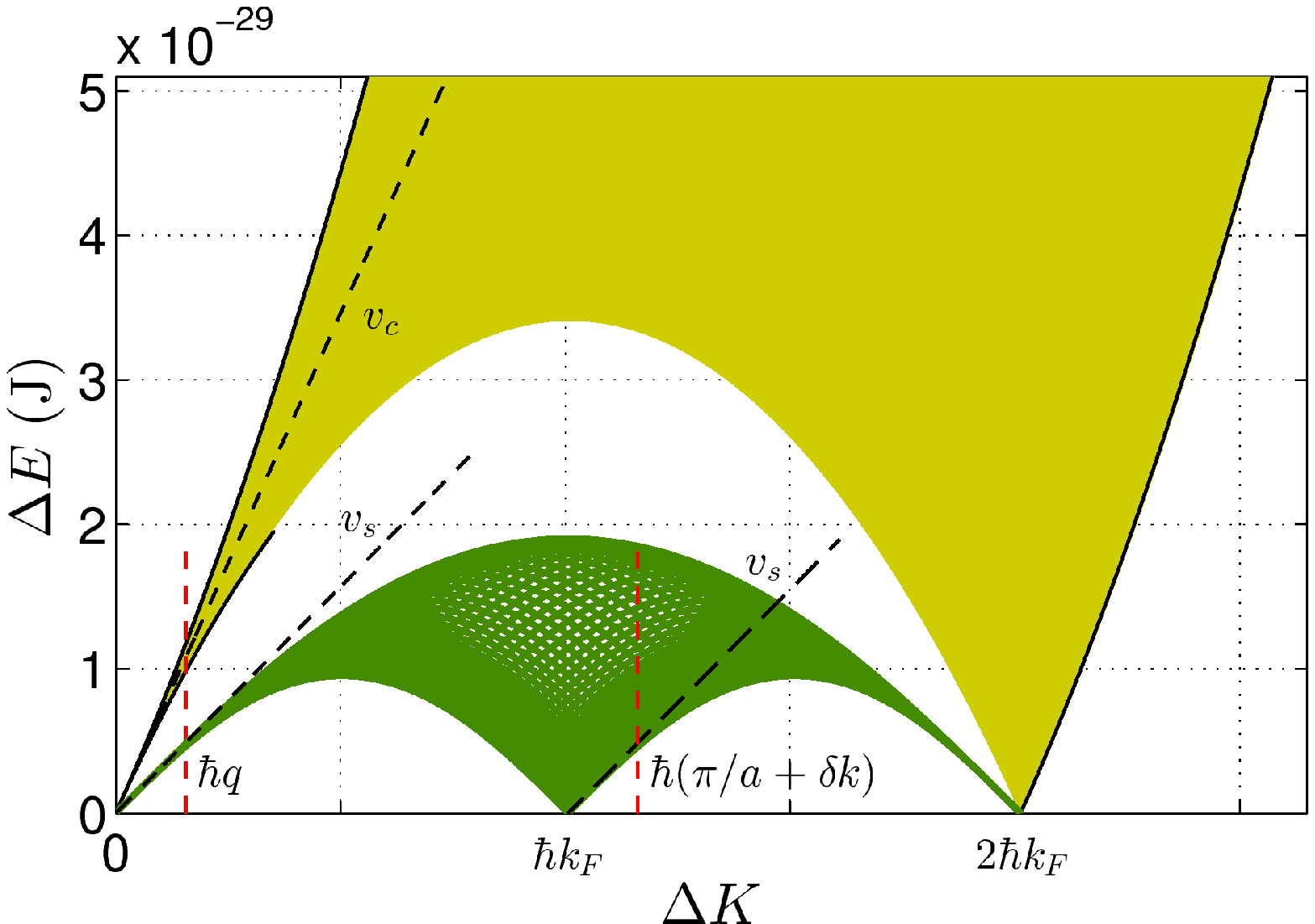}} 	
	{\includegraphics[width=3.4in]{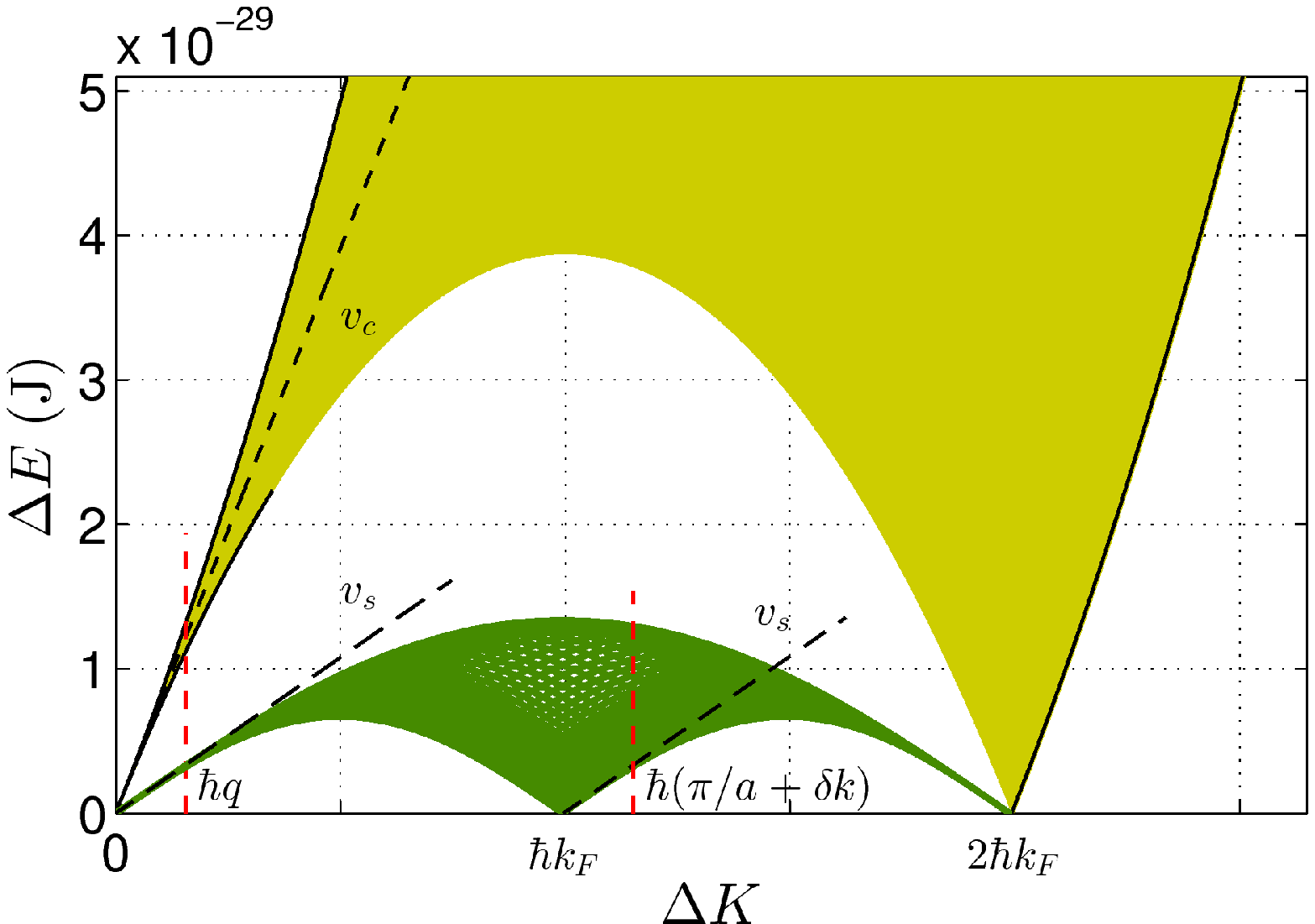}} 		
	\caption{{\small Left panel:  Exact low energy excitation spectra for charge (yellow green) and spin (dark green) at $\gamma = 5.03 \;(a_s=700a_0)$ with the Fermi surface $k_F=n\pi$, density $n=N/L=3 \times 10^6 \;(\rm 1/m)$, $\Delta E=\hbar \omega$. The yellow green  spectrum shows the particle-hole continuum excitation.  The black solid lines indicate  the thresholds of particle-hole excitation which remarkably manifest  the  free fermion-like  dispersion  (\ref{dispersion}) with  an  effective mass $m^* \approx 1.27 m$ at low energy (see Fig.~\ref{effectivemass}). The black dished line in the charge excitation stands for the  charge velocity $v_c$. The dark green spectrum shows the two-spinon excitation, where the black dished  lines stand for  the spin velocities $v_s$ near $\Delta K=0$ and $\hbar k_F$, respectively. The two red dished  lines indicates  the positions of excitation  momenta in charge and spin sectors,  which  are used in the main text. Here we set up  $\Delta K= \hbar q$, $\Delta K= \hbar (\pi a + \delta k)$, $q=\delta k =1.47\;\mu m^{-1}$  for both charge and spin DSFs, respectively. 
	Right panel: Exact low energy excitation spectra for charge (yellow green) and spin (dark green) at $\gamma = 10$, where the effective mass  $m^* \approx 1.22 m$.  We can clearly see that the band structures  are the same for both $ \gamma =5.03$ and $\gamma =10$ except the charges of their  velocities and effective masses. } }
	\label{spectrum}
\end{figure}

At finite temperatures, the imaginary part of the dynamic charge susceptibility  becomes
\begin{eqnarray}
\label{finiteTDSF}
{\rm Im} \chi (q, \omega,,k_F,T,N)=\frac{N  \omega}{2 \hbar^2 q k_F} \pi (n_{q_{-}}-n_{q_{+}})
\end{eqnarray}  
with 
\begin{eqnarray}
\label{fermidistribution}
q_{\pm}=\frac{\omega m^*}{\hbar q}\pm \frac{q}{2},\quad n_q=\frac{1}{{\rm e}^{\beta(\varepsilon_q -\mu)}+1}, \quad \varepsilon_q=\frac{\hbar^2 q^2}{2 m^*}. \label{n-pm}
\end{eqnarray}
Here we demonstrate that this result holds true at low energy not only for weak interaction (as being demonstrated  in \cite{Hulet:2018}), but also for arbitrary interaction strength.

In fact, the result (\ref{LDSorigin}) can be adapted to treat the DSF of the charge for  the interacting Fermi gases (\ref{Ham}).
 The excitation in charge sector display  a similar dispersion structure for both weakly and strongly interacting fermions in the long wave limit. In Fig.~\ref{spectrum}, we show the low-lying excitation for both the charge and the spin obtained by numerically solving the TBA equations. 
From this figure, we observe that the charge excitation can be described by 
\begin{equation}
\omega (q)=v_c |q| \pm \frac{\hbar q^2}{2 m^*}+\cdots, \label{dispersion} 
\end{equation}
where the effective mass $m^*$ can be calculated from the excitation spectrum with the help of the  Bethe ansatz equations  (\ref{BA1}) and (\ref{BA2}), a more detailed calculation will be presented in \cite{Long-Paper}. 
On the other hand, from the TBA equations  (\ref{wholeTBA}), we may evaluate the charge particle-hole excitation  (\ref{dispersion}). 
The TBA equations (\ref{wholeTBA}) at $T=0$ and $H=0$ read
	\begin{eqnarray}
	\label{TBAhzero}
	\varepsilon_c^0 (k) &=&k^2 -\mu + \int_{-\infty}^{\infty} a_1 (k-\lambda)\phi_s^0 (\lambda) {\rm d} \lambda, \nonumber\\
	\phi_s^0 (\lambda) &=&\int_{-k_0}^{k_0} s(\lambda-k) \varepsilon_c^0 (k) {\rm d}k. 
	\end{eqnarray}
namely,
\begin{eqnarray}
\varepsilon_c (k) =k^2 -\mu + \int_{-\infty}^{\infty} \int_{-k_0}^{k_0}a_1 (k-\lambda)s(\lambda-k') \varepsilon_c (k') {\rm d}k' {\rm d} \lambda.
\end{eqnarray}
where we have neglected superscript in $\varepsilon_c^0 (k)$ for simplicity.

 For a particle-hole excitation near the Fermi point $k_0$, the momentum and energy are given by 
\begin{eqnarray}
	K&=&2\pi \int_{0}^{k_0+\Delta k} \rho_c (k) {\rm d}k \\
	\Delta E&=&|\varepsilon_c(k_0+\Delta k)|,
\end{eqnarray}
 where we take $\Delta k <0$ and very small. 
After expanding with $\Delta k$, we obtain 
\begin{eqnarray}
	K&=&2\pi \int_{0}^{k_0} \rho_c(k)  {\rm d}k + 2\pi \rho_c(k_0) \Delta k +2\pi \frac{\rho_c'(k_0)}{2} (\Delta k)^2, \\
	\Delta E&=&|\varepsilon_c(k_0+\Delta k)|=\varepsilon_c(k_0)+\varepsilon_c'(k_0)|\Delta k|+\frac{\varepsilon_c''(k_0)}{2}(\Delta k)^2,
\end{eqnarray}
where $\varepsilon_c(k_0)=0$ by definition. Then we have the total momentum and excitation energy 
\begin{eqnarray}
\Delta K&= &2\pi \rho_c(k_0) \Delta k +2\pi \frac{\rho_c'(k_0)}{2} (\Delta k)^2 \\
\Delta E&=&\varepsilon_c'(k_0)|\Delta k| +\frac{\varepsilon_c''(k_0)}{2}(\Delta k)^2
\end{eqnarray}
After some algebra, we have
\begin{eqnarray}
\Delta k&=&\frac{1}{2\pi \rho_c(k_0)}\Delta K -\frac{\pi \rho_c'(k_0)}{(2 \pi \rho_c(k_0))^3}(\Delta K)^2,\\
\Delta E&=&v_c |\Delta K| + \frac{1}{2 m^*}(\Delta K)^2
\end{eqnarray}  
with 
\begin{eqnarray}
   v_c &=& \frac{\varepsilon_c'(k_0)}{2\pi \rho_c(k_0)}, \\
	\frac{1}{2 m^*} &=&\frac{\varepsilon_c''(k_0)}{2(2 \pi \rho_c(k_0)^2} - \frac{\pi \rho_c'(k_0) \varepsilon_c'(k_0)}{(2 \pi \rho_c(k_0))^3}.
\end{eqnarray}
For strong coupling limit,  after a tedious calculation, we can get the charge velocity and the  effective mass 
\begin{eqnarray}
\label{EMstrong}
v_c \approx 2 \pi n_c \left(  1 - \frac{4 \ln 2}{\gamma} \right), \quad m^*=m\left(1+ \frac{4\ln2}{\gamma}\right).
\end{eqnarray}

For arbitrary interaction strength, the relation of the effective mass versus dimensionless parameter $\gamma=c/n$ can be numerically calculated from the TBA equations (\ref{TBAhzero}), see Fig.~\ref{effectivemass}, where the effective mass ratio $m^*/m \rightarrow 1$ for large $\gamma$. This indicates  that the repulsive Fermi system becomes a real free Fermi system in the strong interaction limitation limit. 
In Fig. 4 of the main text, the corresponding effective masses are $m^*=1.23m,\, 1.255m,\, 1.265m,\, 1.27m$ for $a_s =$ 400$a_0$, 500$a_0$, 600$a_0$, and 700$a_0$, respectively. 

\begin{figure}[tbp]
	\centering
	{\includegraphics[width=3.5in]{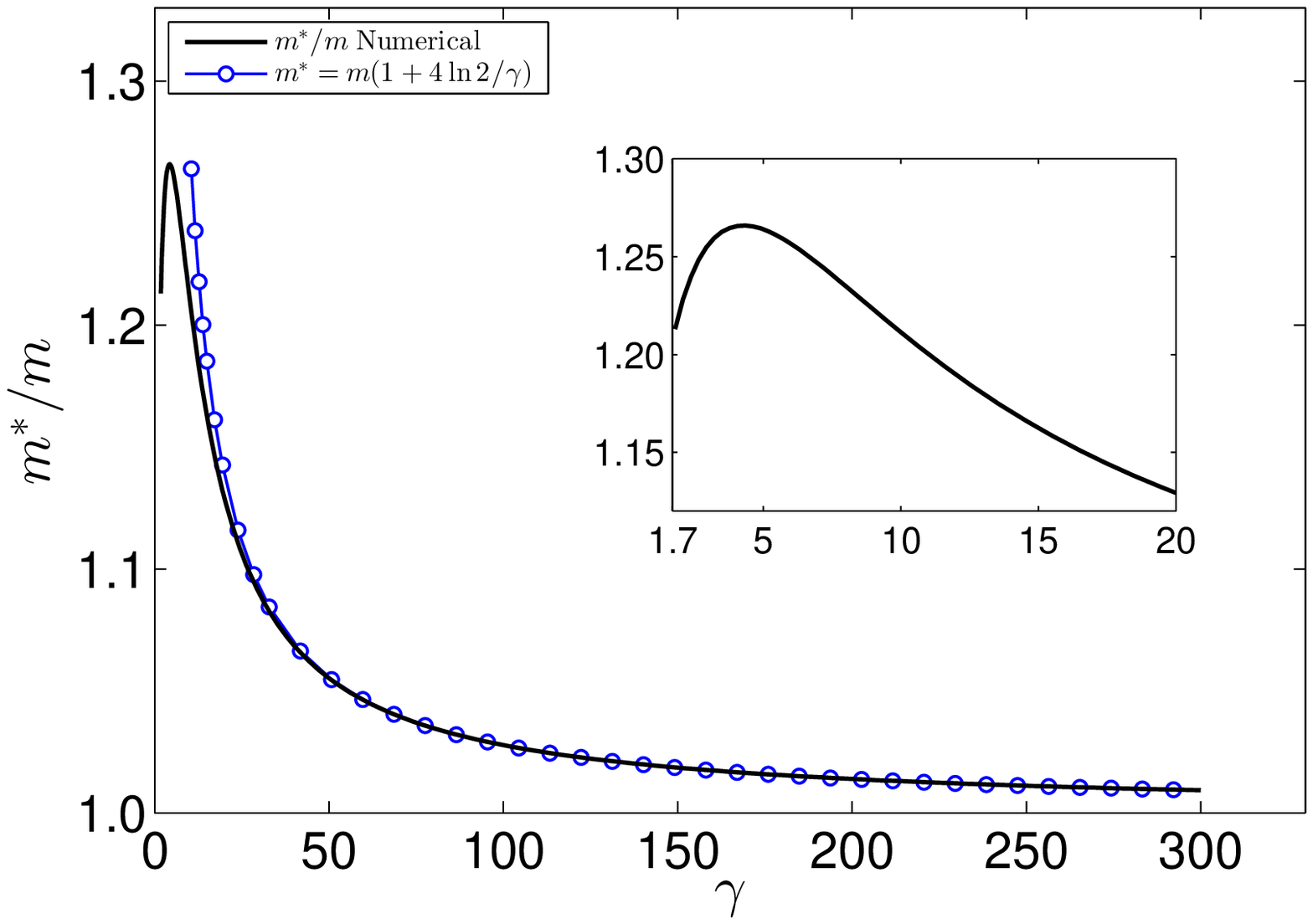}} 		
	\caption{{\small The ratio of $m^*/m$ versus the dimensionless parameter $\gamma$  for the homogeneous repulsive Fermi gas. Here $m^*$ and $m$  denote the  effective mass and the  bare mass of particles. Using  the TBA equations (\ref{TBAhzero}), we  numerically fit  the excitation spectrum within the momentum $\Delta K =[0,\hbar k_F/4]$ according to (\ref{dispersion}) for  different interaction strengths(black solid lines). Analytical effective mass relation (\ref{EMstrong}) is shown with blue-cycle line from interaction strength $\gamma=15$ to $\gamma=300$.  The inset zooms in the effective mass for  interaction strength up to $\gamma =20 $. }}
	\label{effectivemass}
\end{figure}

We observe that the low-energy excitations can be well captured by the leading order in (\ref{dispersion}). 
Fig.~\ref{spectrum} obviously confirms the validity of the DSF Eq.~(\ref{LDSorigin}) at small momentum transfer for the Fermi gases with an arbitrary interaction strength. 
{\em This is mainly because the second term in (\ref{dispersion}) is irrelevant at  low energy. }
For finite repulsion, the Fermi point changes as a function of interaction strength.  Based on the low-lying excitations of the Fermi gas Eq.~(\ref{dispersion}), one can replace the non-interacting Fermi point $k_F$ by the  sound velocities at different interaction strength, i.e. 
\begin{eqnarray}
k_F \rightarrow k_c, \quad k_c=\frac{m^*}{\hbar} v_c.
\end{eqnarray}
Thus the effective charge DSF for interacting Fermi gas  is given by 
\begin{eqnarray}
\label{chargeDSFfinal}
S(q,\omega)=\frac{{\rm Im} \chi (q, \omega,k_c,T,N)}{\pi (1-{\rm e}^{-\beta \hbar \omega})}.
\end{eqnarray}
Here by definition (\ref{vcvsdefination}), the sound velocity $v_c$ can be calculated by zero temperature TBA equations (\ref{TBAhzero}).
From the distribution function $n_q=\frac{1}{{\rm e}^{\beta(\varepsilon_q -\mu)}+1}$ with $\varepsilon_q=\frac{\hbar^2 q^2}{2 m^*}$ and $q_{\pm}=\frac{\omega m^*}{\hbar q}\pm \frac{q}{2}$ and chemical potential  $\mu=\frac{\hbar ^2}{2m^*}k_c^2$, see (\ref{n-pm}), we finally have the  charge DSF 
\begin{eqnarray}
S(q,\omega)&=&\frac{N  \omega / (2 \hbar^2 q k_c)}{ (1-{\rm e}^{-\beta \hbar \omega})} (n_{q_{-}}-n_{q_{+}})\nonumber \\
&=&\frac{N  \omega / (2 \hbar^2 q k_c)}{ (1-{\rm e}^{-\beta \hbar \omega})}  \left\{ \frac{1}{{\rm e}^{\beta \left[ \frac{m^*}{2q^2} \left( \omega -\frac{\hbar q^2}{2 m^*} \right)^2  -
\frac{m^*}{2}v_c^2 \right]}+1}- \frac{1}{{\rm e}^{\beta \left[ \frac{m^*}{2q^2} \left( \omega +\frac{\hbar q^2}{2 m^*} \right)^2  -
\frac{m^*}{2}v_c^2 \right]}+1}  \right\}.
\end{eqnarray}
According to (\ref{dispersion}), we can see that DSF $S(q,\omega)=0$ when $\omega> \omega_{+} =v_c |q| +\frac{\hbar q^2}{2 m^*}$ and $\omega < \omega_{-} =v_c |q|   -\frac{\hbar q^2}{2 m^*}$ at $T \rightarrow 0$, thereby the non-zero width of charge DSF  in Fig.~4 of our main text gives rise to  the width of charge excitation spectrum, i.e., $\Delta \omega =\frac{\hbar q^2}{m^*}$. The maximum value of DSF $S(q,\omega)$ appears at $\omega =v_c q$, which is not effected by the effective mass at low energy.
This shows an effective way of determining the charge velocity from the DSF.

\subsubsection{Spin dynamic structure factor}~

The DSF of spin sector is attributed to many excited states, in which the most important excited states are the two-spinon excitations in $\langle {\rm G}|S^-S^+|{\rm G}\rangle=\sum_{\rm E} \langle {\rm G}|S^-|{\rm E}  \rangle \langle {\rm E}| S^+|{\rm G}\rangle$ when the magnetic field is zero. The only non-zero matrix element in  $\langle {\rm G}|S^-S^+|{\rm G}\rangle$ is contributed from the state with $M^z=1$ as one spin flips up from ground state. 
The logarithm of BAE (\ref{BA1}) and (\ref{BA2}) with string hypothesis (\ref{string}) reduce to the following two sets of  Bethe ansatz equations with the quantum number $\{I_j\}$ and $\{J_{\alpha}^n \}$
\begin{eqnarray}
&&k_jL=2\pi I_j - \sum_{n=1}^{\infty} \sum_{\alpha=1}^{M_n} \theta \left( \frac{2(k_j-\lambda_{\alpha}^n)}{nc} \right), \quad j=1,2,\cdots,N, \\
&&\sum_{j=1}^{N} \theta \left( \frac{2(k_j-\lambda_{\alpha}^n)}{nc} \right) =2\pi J_{\alpha}^{n} + \sum_{m=1}^{\infty} \sum_{\beta=1}^{M_m} \Theta_{mn} \left( \frac{2(\lambda_{\alpha}^m-\lambda_{\beta}^m)}{c} \right),
\end{eqnarray}
where  $ \alpha =1,2,\cdots,M_n, \; n\geq 1$ and  $M_n$ is the number of length-$n$ string, $\theta(x)=2 \tan^{-1} (x)$, and $\Theta_{mn}(x)$ is defined by
\begin{eqnarray}
\Theta_{mn}(x) =  \left \{
\begin{array}{rcl}
&&\theta\left( \frac{x}{|n-m|} \right)+2\theta\left( \frac{x}{|n-m|+2} \right)+ \cdots +2\theta\left( \frac{x}{n+m-2} \right)+\theta\left( \frac{x}{m+n} \right) \quad \text{for} \quad n \neq m,  \\
&&2\theta\left( \frac{x}{2} \right)+2\theta\left( \frac{x}{4} \right)+ \cdots +2\theta\left( \frac{x}{2n-2} \right)+\theta\left( \frac{x}{2n} \right) \quad \text{for} \quad n = m.
\end{array}
\right.
\end{eqnarray}
The quantum number $I_j$ for charge  take  distinct integers (or half-odd integers) for even (odd) $\sum_\alpha  M_\alpha $, explicitly 
\begin{equation}
 I_j\in \sum_{n=1}^{\infty} \frac{M_n}{2} + \mathbb{Z}.
\end{equation}
The spin quantum number $J_{\alpha}^n$ are distinct integers (half-odd integers) for odd (even) $N-M_m$, which satisfy
\begin{eqnarray}
&& J_{\alpha}^n \in \frac{N-M_n}{2}+\frac{1}{2} + \mathbb{Z}, \\
&&|J_{\alpha}^n | \leq I_{+}^n= \frac{N}{2} -\sum_{m=1}^{n}mM_m - n \sum_{m=n+1}^{\infty} M_m +\frac{M_n}{2} -\frac{1}{2}, \label{jmax}\\
&& J_{\alpha}^n = -I_{+}^n, -I_{+}^n+1, -I_{+}^n+2,\cdots, I_{+}^n-1,I_{+}^n .
\end{eqnarray}
The total momentum of the system is 
\begin{equation}
\label{totalmomentum}
K=\sum_{j=1}^{N}k_j = \frac{2\pi}{L} \left( \sum_{j=1}^{N} I_j + \sum_{\alpha=1}^{M_n} \sum_{n=1}^{\infty} J_{\alpha}^n  \right).
\end{equation}
Without losing generality, we consider the excited state with one spin flip up from the  ground state with $N=4\mathbb{Z} $, namely
$M=M_1=N/2-1$, without high length strings, i.e., $M_n=0$, $ n \geq 2$. Thus we obtain the total excited momenta
\begin{equation}
\label{spinexcitationmomentum}
\Delta K_{\rm spinon} =K-K_{\rm G}=n\pi- 2\pi \sum_{j=1}^{2} \int_{0}^{\lambda_j^h} \rho_s^0(\lambda) {\rm d}\lambda,\quad{\text{or}}\quad \Delta K_{\rm spinon} =-n\pi- 2\pi \sum_{j=1}^{2} \int_{0}^{\lambda_j^h} \rho_s^0(\lambda) {\rm d}\lambda
\end{equation}
presenting microscopic origin of the two deconfined spions.  
Next, we derive the energy of two-spionon excitation, where the model  exhibits two deconfined two hole quasimomenta density functions in low-energy spin excitations.

We first define $\bar{\rho}_s(\lambda)=\rho_s (\lambda)+\rho_s^{\rm h} (\lambda)= \rho_s (\lambda)+\frac{1}{L}\sum_{j=1}^{2} \delta (\lambda-\lambda_j^{\rm h})$.  From the densities functions (\ref{zerodensitycharge}) and (\ref{zerodensityspin}), we have
\begin{align}
\rho_c(k) &=\frac{1}{2 \pi}+ \int_{\lambda_-}^{\lambda_+}a_1 (k-\lambda) \bar{\rho}_s(\lambda) {\rm d}\lambda -\frac{1}{L}\sum_{j=1}^{2} a_1 (\lambda-\lambda_j^{\rm h}),\\
\bar{\rho}_s (\lambda)&=\int_{k_-}^{k_+}  a_1 (\lambda-k) \rho_c (k) {\rm d}k - \int_{\lambda_-}^{\lambda_+} a_2 (\lambda-\lambda') \bar{\rho}_s (\lambda') {\rm d} \lambda +\frac{1}{L}\sum_{j=1}^{2} a_2 (\lambda-\lambda_j^{\rm h}),
\end{align}
After some algebra, we get the excitation energy 
\begin{equation}
\label{excitationspin}
\Delta E_{\rm spinon}= \int_{k_-}^{k_+} \rho_c(k) k^2 {\rm d}k - \int_{-k_0}^{k_0} \rho_c^0(k) k^2 {\rm d}k =\int_{-k_0}^{k_0} \Delta \rho_c(k) (k^2-\mu) {\rm d}k+2(k_0^2-\mu)  \rho_c(k_0) \Delta k.
\end{equation}
We further calculate the term 
\begin{eqnarray}
\int_{-k_0}^{k_0} \Delta \rho_c (k)(k^2-\mu) {\rm d}k &=& -2(k_0^2-\mu)\rho_c(k_0)\Delta k
-\int_{-k_0}^{k_0} \frac{1}{L}\sum_{j=1}^{2} a_1 (k-\lambda_j^{\rm h}) \varepsilon_c^0(k) {\rm d} k \nonumber\\
&& +\int_{-\lambda_0}^{\lambda_0} \frac{1}{L}\sum_{j=1}^{2} a_2 (\lambda-\lambda_j^{\rm h}) \phi_s^0(\lambda) {\rm d} \lambda.
\end{eqnarray}
With the help of the TBA equations  (\ref{TBAhzero}), we obtained the energy of two-spinon excitation
\begin{align}
\label{spinexcitationenergy}
\Delta E_{\rm spinon} =-\sum_{j=1}^{2} \left[ \int_{-k_0}^{k_0} a_1 (k-\lambda_j^{\rm h}) \varepsilon_c^0(k) {\rm d} k - \int_{-\lambda_0}^{\lambda_0}  a_2 (\lambda-\lambda_j^{\rm h}) \phi_s^0(\lambda) {\rm d} \lambda \right] =-\sum_{j=1}^{2}\phi_s^0 (\lambda_j^{\rm h}),
\end{align}

The low-energy excitation in the spin sector is displayed in Fig.~\ref{spectrum}, which are obtained from solving the TBA equations in terms of the equations  (\ref{spinexcitationenergy}),(\ref{spinexcitationmomentum}). This two-spinon excitation spectrum holds for the whole interaction regime. However, when the interaction increase, the spin excitation band becomes lower, and vanishes in the limit $\gamma \to \infty$. 

It turns out that the effective Heisenberg chain essentially capture magnetic ordering and fractional excitations.
The two-spinon continuum spectra shown in Fig.~\ref{spectrum} is a common feature of the Fermi gas with arbitrary interaction. 
For arbitrary interaction and at low temperatures, the spin DSF of Luttinger liquid is given in general by \cite{giamarch:book}
\begin{eqnarray}
S(q,\omega, T)\propto T^{\frac{1-4K}{2K}}\left( \frac{1}{1-e^{-\frac{\hbar \omega}{k_B T }} } \right){\rm Im}\left\{
 \frac{\Gamma \left(\frac{1}{8K} -\mathrm{i} \hbar \frac{v_sq/a+\omega }{4\pi k_BT}  \right) \Gamma \left(\frac{1}{8K} -\mathrm{i} \hbar \frac{v_s q/a-\omega }{4\pi k_BT}  \right)}{ \Gamma \left(1-\frac{1}{8K} -\mathrm{i} \hbar \frac{v_s q/a+\omega }{4\pi k_BT}  \right)\Gamma \left(1-\frac{1}{8K} -\mathrm{i} \hbar \frac{v_s q/a-\omega }{4\pi k_BT}  \right)} 
 \right\}, \label{Luttinger-DSF-S} 
\end{eqnarray} 
where $ a$ is the spin  lattice constant, $K$ is the Luttinger parameter ($K=1/2$ when magnetic field $H=0$) and $v_s$ is the spin velocity. This formula holds true for the two-spinon excitations with the wave vector $k=0,\pi$. 
Based on this analysis, we see that the spin DSF of spin-$1/2$ repulsive Fermi gas in the spin charge separated regime can be approximated by that of a Heisenberg spin-$1/2$ chain.

\begin{figure}[tbp]
	\centering
	{\includegraphics[width=5.7in]{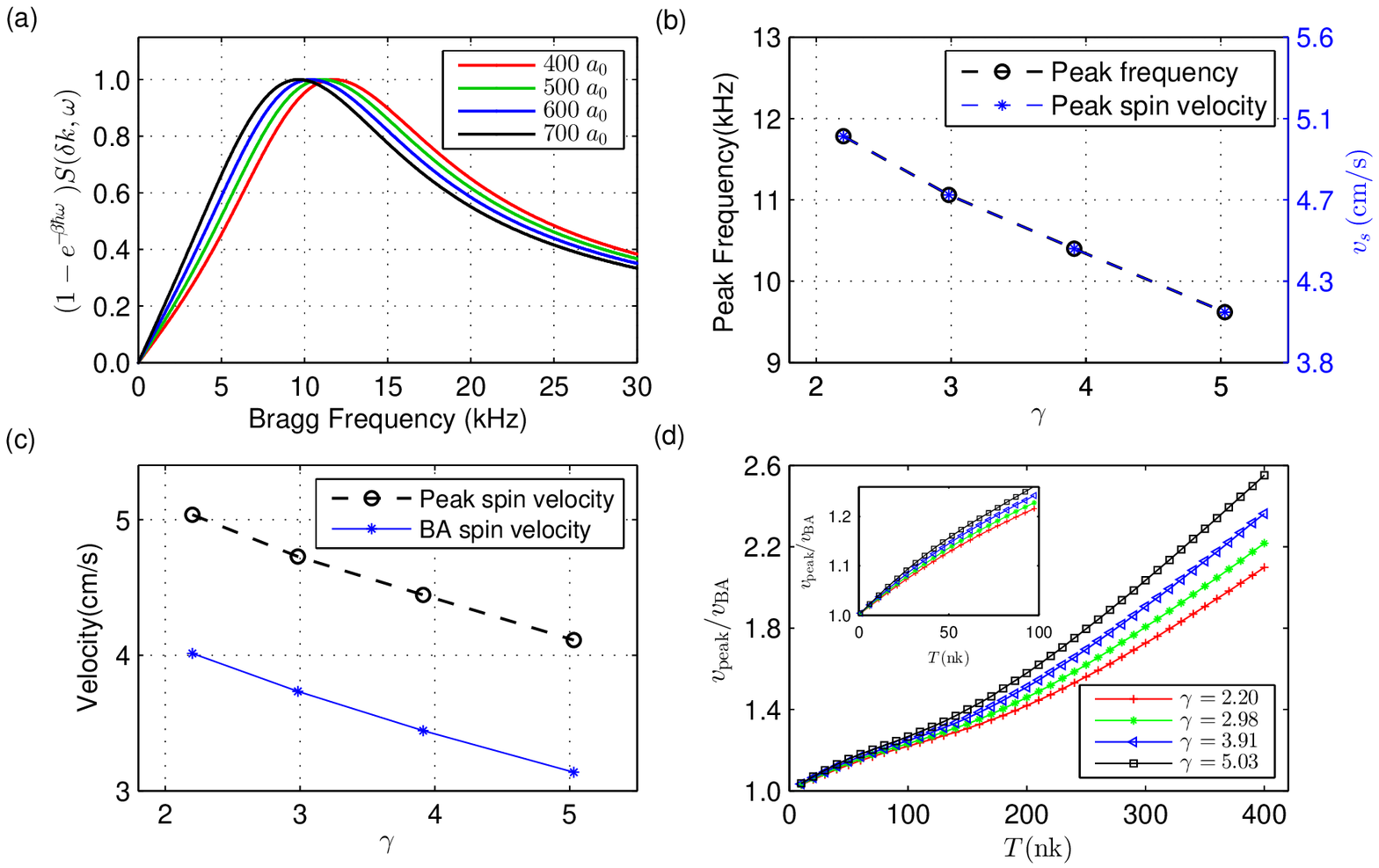}} 		
	\caption{{\small (a) Normalized spin DSFs of a homogeneous tube with several different values of interaction strengths at $T=200 \;{\rm nk}$. (b) Peak frequency (left vertical axis) of each spectrum vs. the effective interaction strength $\gamma=c/n$, the right vertical axis indicates the peak spin velocity defined as the ratio of the peak frequency and the momentum transfer $\delta k$. (c) The peak spin velocities and the BA spin velocities, and the BA spin velocities are used in the expression  (\ref{TLLDSFSLL}) and obtained by solving TBA equations. (d) The ratio of the peak velocity over BA velocity $v_{\rm peak}/v_{\rm BA}$ vs. temperature $T$ at different interaction strengths. The inset zooms in the ratio at low temperatures.}}
	\label{homospin}
\end{figure}

The linear dispersion of spinons in long wave limit can be well discriebed by Tommanaga-Luttinger liquid (TLL) theory. At finite temperatures, near the wave vector $k=0,\pi$, the spin DSF in the spin chain was obtained by the TLL theory \cite{caux:2013}. Explicitly, from (\ref{Luttinger-DSF-S}), around $k=\pi/a+\delta k$ with the  the lattice constant $a=L/N$ and $H=0$,  it is given by 
\begin{eqnarray}
S(\delta k, \omega)&=& \frac{1}{1-{\rm e}^{- \beta \hbar \omega}}\frac{A_{LL}}{k_B T} {\rm Im} \left[ \rho \left( \frac{\hbar \omega+v_s \hbar \delta k}{4 \pi k_B T}\right)  \rho \left( \frac{\hbar \omega-v_s \hbar \delta k}{4 \pi k_B T}\right) \right],\label{TLLDSFSLL}
\end{eqnarray}
where $\rho(x)= \Gamma(1/4-ix)/\Gamma(3/4-ix)$, $v_s=(\pi/2)J$ for spin chain.  $A_{LL}=-c_{\perp}^2 \alpha /2$ is a constant with the length scale parameter $\alpha$ and constant factor $c_{\perp}$. It is worth noting that  this form of DSF is valid only for the linear dispersion of spinons.

Fig.~\ref{homospin}(a) plots the spin dynamic structure factors (DSFs) versus Bragg frequency, which is reproduced from Fig.~4(c) of the main text. In converting from the dimensionless units to real units, we have assumed a system of spin-balanced $^6$Li atoms with total atom number $N=60$, confined in a homogeneous tube with length $L=20\,\mu$m, and transverse harmonic trap with trapping potential $\omega_\perp = (2\pi)\,198$kHz, at temperature $T=120$nK. The momentum transfer for calculating the DSF is taken to be $\delta k = 1.47 \,(\mu {\rm m})^{-1}$ In the figure, we have included curves corresponding to 4 different values of interaction strength characterized by the 3D scattering length as 400$a_0$, 500$a_0$, 600$a_0$, and 700$a_0$, which correspond to a dimensionless effective interaction strength $\gamma=c/n=$2.20, 2.98, 3.91, and 5.03, respectively. Fig.~\ref{homospin}(b) shows the peak frequencies and peak velocities which are read off from the charge  DSFs' peak positions as functions of $\gamma$. Here the peak velocity is define as peak frequency/$\delta k$. Unlike in the case for the charge DSF, where the peak velocity is nearly identical to the charge sound velocity, here the spin peak velocity is different from the spin sound velocity as shown in Fig.~\ref{homospin}(c). However, both velocities exhibit a similar dependence on $\gamma$. In Fig.~\ref{homospin}(d), we plot the ratio of the peak velocity and the spin sound velocity as a function of temperature. One can see that, this ratio tends to 1 at $T=0$ and increases as $T$ increases. Such a plot can help experimentalists to extract the value of the spin sound velocity from the measured spin DSF.

\end{widetext} 

\begin{thebibliography}{10}

\bibitem{Landau:2008}
Gordon Baym and Christopher Pethick.
\newblock {\em Landau Fermi-liquid theory: concepts and applications}.
\newblock John Wiley \& Sons, 2008.

\bibitem{quantumliquid:2018}
David Pines.
\newblock {\em Theory of Quantum Liquids: Normal Fermi Liquids}.
\newblock CRC Press, 2018.

\bibitem{schollwock:2011density}
Ulrich Schollw{\"o}ck.
\newblock The density-matrix renormalization group: a short introduction.
\newblock {\em Philosophical Transactions of the Royal Society A: Mathematical,
  Physical and Engineering Sciences}, 369(1946):2643--2661, 2011.
  
  

\bibitem{kollath2005spin}
C~Kollath, U~Schollw{\"o}ck, and W~Zwerger.
\newblock Spin-charge separation in cold fermi gases: A real time analysis.
\newblock {\em Physical review letters}, 95(17):176401, 2005.



\bibitem{green:book}
Eleftherios~N Economou.
\newblock {\em Green's functions in quantum physics}, volume~7.
\newblock Springer Science \& Business Media, 2006.

\bibitem{haldane:1981}
F. D. M. Haldane.
\newblock `Luttinger liquid theory' of one-dimensional quantum fluids. i.
  properties of the Luttinger model and their extension to the general 1D
  interacting spinless fermi gas.
\newblock {\em Journal of Physics C: Solid State Physics}, 14(19):2585, 1981.

\bibitem{Giamarchi:book}
T~Giamarchi.
\newblock Quantum physics in one dimension oxford science publications.
\newblock {\em New York}, 2004.

\bibitem{Imambekov:2012}
Adilet Imambekov, Thomas~L Schmidt, and Leonid~I Glazman.
\newblock One-dimensional quantum liquids: Beyond the luttinger liquid
  paradigm.
\newblock {\em Reviews of Modern Physics}, 84(3):1253, 2012.

\bibitem{Recati:PhysRevLett.90.020401}
A.~Recati, P.~O. Fedichev, W.~Zwerger, and P.~Zoller.
\newblock Spin-charge separation in ultracold quantum gases.
\newblock {\em Physical Review Letters}, 90:020401, 2003.

\bibitem{Guan:2012}
JY~Lee, Xi-Wen Guan, Kazumitsu Sakai, and MT~Batchelor.
\newblock Thermodynamics, spin-charge separation, and correlation functions of
  spin-1/2 fermions with repulsive interaction.
\newblock {\em Physical Review B}, 85(8):085414, 2012.

\bibitem{mestyan2019spin}
M{\'a}rton Mesty{\'a}n, Bruno Bertini, Lorenzo Piroli, and Pasquale Calabrese.
\newblock Spin-charge separation effects in the low-temperature transport of
  one-dimensional fermi gases.
\newblock {\em Physical Review B}, 99(1):014305, 2019.

\bibitem{PhysRevB.101.035149}
Ovidiu~I. P\^a\ifmmode~\mbox{\c{t}}\else \c{t}\fi{}u, Andreas Kl\"umper, and
  Angela Foerster.
\newblock Quantum critical behavior and thermodynamics of the repulsive
  one-dimensional hubbard model in a magnetic field.
\newblock {\em Physical Review B}, 101:035149, 2020.

\bibitem{Sachdev_2001}
S.~Sachdev.
\newblock {\em {Quantum Phase Transitions}}.
\newblock Cambridge University Press, Cambridge, 2001.

\bibitem{Guan:RMP}
Xi-Wen Guan, Murray~T Batchelor, and Chaohong Lee.
\newblock Fermi gases in one dimension: From Bethe ansatz to experiments.
\newblock {\em Reviews of Modern Physics}, 85(4):1633, 2013.

\bibitem{Kinoshita:2004}
Toshiya Kinoshita, Trevor Wenger, and David~S Weiss.
\newblock Observation of a one-dimensional tonks-girardeau gas.
\newblock {\em Science}, 305(5687):1125--1128, 2004.

\bibitem{Kinoshita:2006}
Toshiya Kinoshita, Trevor Wenger, and David~S Weiss.
\newblock A quantum newton's cradle.
\newblock {\em Nature}, 440(7086):900, 2006.

\bibitem{Paredes:2004}
Bel{\'e}n Paredes, Artur Widera, Valentin Murg, Olaf Mandel, Simon F{\"o}lling,
  Ignacio Cirac, Gora~V Shlyapnikov, Theodor~W H{\"a}nsch, and Immanuel Bloch.
\newblock Tonks--girardeau gas of ultracold atoms in an optical lattice.
\newblock {\em Nature}, 429(6989):277--281, 2004.

\bibitem{Haller:2009}
Elmar Haller, Mattias Gustavsson, Manfred~J Mark, Johann~G Danzl, Russell Hart,
  Guido Pupillo, and Hanns-Christoph N{\"a}gerl.
\newblock Realization of an excited, strongly correlated quantum gas phase.
\newblock {\em Science}, 325(5945):1224--1227, 2009.

\bibitem{Pagano:2014}
Guido Pagano, Marco Mancini, Giacomo Cappellini, Pietro Lombardi, Florian
  Sch{\"a}fer, Hui Hu, Xia-Ji Liu, Jacopo Catani, Carlo Sias, Massimo Inguscio,
  et~al.
\newblock A one-dimensional liquid of fermions with tunable spin.
\newblock {\em Nature Physics}, 10(3):198--201, 2014.

\bibitem{liao:2010}
Yean-an Liao, Ann Sophie~C Rittner, Tobias Paprotta, Wenhui Li, Guthrie~B
  Partridge, Randall~G Hulet, Stefan~K Baur, and Erich~J Mueller.
\newblock Spin-imbalance in a one-dimensional fermi gas.
\newblock {\em Nature}, 467(7315):567, 2010.

\bibitem{Yang:2017}
Bing Yang, Yang-Yang Chen, Yong-Guang Zheng, Hui Sun, Han-Ning Dai, Xi-Wen
  Guan, Zhen-Sheng Yuan, and Jian-Wei Pan.
\newblock Quantum criticality and the tomonaga-luttinger liquid in
  one-dimensional bose gases.
\newblock {\em Physical review letters}, 119(16):165701, 2017.

\bibitem{Hulet:2018}
TL~Yang, P~Gri{\v{s}}ins, YT~Chang, ZH~Zhao, CY~Shih, Thierry Giamarchi, and
  RG~Hulet.
\newblock Measurement of the dynamical structure factor of a 1d interacting
  fermi gas.
\newblock {\em Physical review letters}, 121(10):103001, 2018.

\bibitem{PhysRevLett.122.090601}
M.~Schemmer, I.~Bouchoule, B.~Doyon, and J.~Dubail.
\newblock Generalized hydrodynamics on an atom chip.
\newblock {\em Physical review letters}, 122:090601, Mar 2019.

\bibitem{vijayan2020time}
Jayadev Vijayan, Pimonpan Sompet, Guillaume Salomon, Joannis Koepsell, Sarah
  Hirthe, Annabelle Bohrdt, Fabian Grusdt, Immanuel Bloch, and Christian Gross.
\newblock Time-resolved observation of spin-charge deconfinement in fermionic
  Hubbard chains.
\newblock {\em Science}, 367(6474):186--189, 2020.


\bibitem{hilker2017revealing}
Timon~A Hilker, Guillaume Salomon, Fabian Grusdt, Ahmed Omran, Martin Boll,
  Eugene Demler, Immanuel Bloch, and Christian Gross.
\newblock Revealing hidden antiferromagnetic correlations in doped Hubbard
  chains via string correlators.
\newblock {\em Science}, 357(6350):484--487, 2017.

\bibitem{bohrdt2018angle}
Annabelle Bohrdt, D~Greif, E~Demler, M~Knap, and F~Grusdt.
\newblock Angle-resolved photoemission spectroscopy with quantum gas
  microscopes.
\newblock {\em Physical Review B}, 97(12):125117, 2018.


\bibitem{barfknecht2019dynamics}
Rafael~Emilio Barfknecht, Angela Foerster, and Nikolaj~Thomas Zinner.
\newblock Dynamics of spin and density fluctuations in strongly interacting
  few-body systems.
\newblock {\em Scientific reports}, 9(1):1--11, 2019.



\bibitem{Kim:1996}
C~Kim, AY~Matsuura, Z-X Shen, N~Motoyama, H~Eisaki, S~Uchida, Takami Tohyama,
  and S~Maekawa.
\newblock Observation of spin-charge separation in one-dimensional SrCuO${}_2$.
\newblock {\em Physical review letters}, 77(19):4054, 1996.


\bibitem{auslaender2005spin}
OM~Auslaender, H~Steinberg, A~Yacoby, Y~Tserkovnyak, BI~Halperin, KW~Baldwin,
  LN~Pfeiffer, and KW~West.
\newblock Spin-charge separation and localization in one dimension.
\newblock {\em Science}, 308(5718):88--92, 2005.

\bibitem{Yang:1967}
Chen-Ning Yang.
\newblock Some exact results for the many-body problem in one dimension with
  repulsive delta-function interaction.
\newblock {\em Physical Review Letters}, 19(23):1312, 1967.

\bibitem{Gaudin:1967}
M~Gaudin.
\newblock Un systeme a une dimension de fermions en interaction.
\newblock {\em Physics Letters A}, 24(1):55--56, 1967.

\bibitem{Lai:1971}
CK~Lai.
\newblock Thermodynamics of fermions in one dimension with a $\delta$-function
  interaction.
\newblock {\em Physical Review Letters}, 26(24):1472, 1971.

\bibitem{Lai:1973}
CK~Lai.
\newblock Thermodynamics of a one-dimensional system of fermions with a
  repulsive $\delta$-function interaction.
\newblock {\em Physical Review A}, 8(5):2567, 1973.

\bibitem{Takahashi:1971}
Minoru Takahashi.
\newblock One-dimensional electron gas with delta-function interaction at
  finite temperature.
\newblock In {\em Exactly Solvable Models Of Strongly Correlated Electrons},
  pages 388--406. World Scientific, 1994.

\bibitem{Supp}
In this supplementary material, we present basic introduction to the Bethe
  ansatz equations for the 1d spin-$1/2$ fermi gas and partial derivations of
  properties of the spin-charge separated and disrupted liquids.

\bibitem{Wilson_1975}
K.~G. Wilson.
\newblock The renormalization group: critical phenomena and the {Kondo}
  problem.
\newblock {\em Review of Modern Physics}, 47:773--840, 1975.


\bibitem{Guan:2013PRL}
X.-W. Guan, X.-G. Yin, A.~Foerster, M.~T. Batchelor, C.-H. Lee, and H.-Q. Lin.
\newblock Wilson ratio of fermi gases in one dimension.
\newblock {\em Phys. Rev. Lett.}, 111:130401, Sep 2013.


\bibitem{Takahashi_2005}
M. Takahashi, 
\newblock Thermodynamics of one-dimensional solvable models
\newblock Cambridge University Press, 2005.


\bibitem{Long-Paper}
Feng He, Yuzhu Jiang, Hai-Qing Lin, Randall~G Hulet, H~Pu, and Xi-Wen Guan.
\newblock Spin-charge separated and disrupted liquids: A comprehensive study,
  in preparation, 2020.
  
  \bibitem{Fiete:2007}
  Gregory A. Fiete.
  \newblock The spin-incoherent Luttinger liquid. 
  \newblock {\em Review of Modern Physics}, 79(3):801, 2007.
  
  \bibitem{Cheianov:2004} Vadim V. Cheianov and M. B. Zvonarev.
  \newblock Nonunitary Spin-Charge Separation in a One-Dimensional Fermion Gas.
\newblock {\em Phys. Rev. Lett.},  92:176401, April 2004.

\bibitem{Oshanii_PRL_1998}
Maxim Olshanii.
\newblock Atomic scattering in the presence of an external confinement and a
  gas of impenetrable bosons.
\newblock {\em Physical Review Letters}, 81(5):938, 1998.

\bibitem{Yang:PhD}
Tsung-Lin Yang.
\newblock {\em Dynamical Response of an Interacting 1-Dimensional Fermi Gas}.
\newblock PhD thesis, Rice University, 2018.

\bibitem{Hoinka:2012}
S~Hoinka, M~Lingham, M~Delehaye, and CJ~Vale.
\newblock Dynamic spin response of a strongly interacting fermi gas.
\newblock {\em Physical review letters}, 109(5):050403, 2012.

\bibitem{Brunello:2001}
A.~Brunello, F.~Dalfovo, L.~Pitaevskii, S.~Stringari, and F.~Zambelli.
\newblock Momentum transferred to a trapped bose-einstein condensate by
  stimulated light scattering.
\newblock {\em Phys. Rev. A}, 64:063614, Nov 2001.

\bibitem{cherny:2006polarizability}
Alexander~Yu Cherny and Joachim Brand.
\newblock Polarizability and dynamic structure factor of the one-dimensional
  bose gas near the tonks-girardeau limit at finite temperatures.
\newblock {\em Physical Review A}, 73(2):023612, 2006.

\bibitem{Schulz:1991}
HJ~Schulz.
\newblock Correlated fermions in one dimension.
\newblock In {\em Exactly Solvable Models Of Strongly Correlated Electrons},
  pages 198--215. World Scientific, 1994.

\end{thebibliography}

\begin{thebibliography}{99}

\bibitem{Oshanii_PRL_1998} M. Olshanii, Atomic scattering in the presence of an external confinement and a gas of impenetrable bosons. Phys.  Rev.  Lett.,  81(5):938, 1998.

	\bibitem{Takahashi:1999}M. Takahashi,  {\it Thermodynamics of  One-Dimensional Solvable Models}(Cambridge University Press,  Cambridge, 1999).	
	
	\bibitem{Lai:1971}Lai C K. Thermodynamics of fermions in one dimension with a δ-function interaction, Physical Review Letters, 1971, 26(24): 1472.
	
	\bibitem{Lai:1973}Lai C K. Thermodynamics of a one-dimensional system of fermions with a repulsive δ-function interaction, Physical Review A, 1973, 8(5): 2567.
	
	\bibitem{Takahashi:1971} M. Takahashi: One-dimensional electron gas with delta-function interaction at finite temperature Prog. Theor. Phys.1971,26:1388.
	
	\bibitem{Lee:2012} Lee J Y, Guan X W, Sakai K, et al. Thermodynamics, spin-charge separation, and correlation functions of spin-1/2 fermions with repulsive interaction,  Physical Review B, 2012, 85(8): 085414.
	

	\bibitem{caux:2013} Lake B, Tennant D A, Caux J S, et al. Multispinon continua at zero and finite temperature in a near-ideal Heisenberg chain,  Physical review letters, 2013, 111(13): 137205.
	
	
	
\bibitem{Ess05}F. H. L. Essler, H. Frahm, F. G\"{o}hmann, A. Kl\"{u}mper  and V. E. Korepin,
{ The One-Dimensional Hubbard Model} (Cambridge University Press, Cambridge, 2005).



\bibitem{Long-Paper}
Feng He, Yuzhu Jiang, Hai-Qing Lin, Randall~G Hulet, H~Pu, and Xi-Wen Guan.
\newblock Spin-charge separated and disrupted liquids: A comprehensive study,
  in preparation, 2020.
  
  \bibitem{He:2017}He F, Y. Jiang, Y.-C. Yu, H.-Q. Lin and X.-W. Guan. Quantum criticality of spinons. Physical Review B, 2017, 96(22): 220401.

  \bibitem{Cherny:2006}Cherny A Y, Brand J. Polarizability and dynamic structure factor of the one-dimensional Bose gas near the Tonks-Girardeau limit at finite temperatures[J]. Physical Review A, 2006, 73(2): 023612.
 

\bibitem{Hulet:2018}
T L~Yang, P~Gri{\v{s}}ins, Y T~Chang, Z H~Zhao, C Y~Shih, Thierry Giamarchi, and
  RG~Hulet.
\newblock  Measurement of the dynamical structure factor of a 1d interacting
  fermi gas.
\newblock {Physical review letters}, 121(10):103001, 2018.


\bibitem{giamarch:book}
T Giamarchi.
\newblock   Quantum physics in one dimension. New York, 2004.

   \end{thebibliography}
\end{document}